\documentclass[preprint,showpacs,showkeys,aps,prd,amsfonts,eqsecnum,nofootinbib]
{revtex4}
\usepackage{graphicx,epsfig}
\usepackage[dvips]{color}
\newcommand{\no}{\noindent}

\def\gsim{\lower0.5ex\hbox{$\:\buildrel >\over\sim\:$}}
\def\lsim{\lower0.5ex\hbox{$\:\buildrel <\over\sim\:$}}
\begin{document}
\preprint{CUMQ/HEP 142}
%
%
\title{\Large RARE DECAY OF THE TOP $t \rightarrow cl{\bar l}$ AND SINGLE TOP PRODUCTION AT ILC}
\author{Mariana Frank}\email[]{mfrank@alcor.concordia.ca}
\author{Ismail Turan}\email[]{ituran@physics.concordia.ca}
\affiliation{Department of Physics, Concordia University, 7141
Sherbrooke Street West, Montreal, Quebec, CANADA H4B 1R6}
\date{\today}

\begin{abstract}
We perform a complete and detailed analysis of the flavor changing neutral current
rare top quark decays $t \to cl^+l^-$ and $t \to c \nu_i{\bar \nu}_i$ at one-loop level in the Standard Model, Two Higgs Doublet Models (I and II) and in MSSM. The branching ratios are very small in all models, ${\cal O}(10^{-14})$, except for the case of the unconstrained MSSM, where they can reach ${\cal O}(10^{-6})$ for $e^+e^-$ and $\nu_i{\bar \nu}_i$, and ${\cal O}(10^{-5})$ for $\tau^+ \tau^-$. This branching ratio is comparable to the ones for $t \to c V, cH$. We also study the production rates of single top and the forward-backward asymmetry in $e^+e^- \to t{\bar c}$  and comment on the observability of such a signal at the International Linear Collider.

\end{abstract}
\maketitle
\section{Introduction}\label{sec:intro}
\no
With the advent of the CERN Large Hadron Collider (LHC) \cite{LHC},  80 million top quark pairs will be produced per year \cite{aguilar}. This number will increase
by one order of magnitude with the high luminosity option.
Therefore, the properties of top quarks can be examined 
with significant precision at LHC. Thus top quark physics will be a testing ground for new phenomena, be it electroweak symmetry breaking, or the existence of interactions forbidden in the Standard Model (SM). Flavor Changing Neutral Currents (FCNC)
in general, and of the top quark in
particular, play an important role for validation of  the
SM and for New Physics (NP) signals. In the SM, there are no FCNC mediated by the $\gamma, Z, g$ or $H$-boson at tree level, and FCNC induced by radiative effects are highly suppressed  \cite{Eilam:1990zc,Mele:1998ag}. Higher order contributions are proportional to $(m_i^2-m_j^2)/M_W^2$, with $m_i,m_j$ the masses of the quarks in the loop diagrams and $M_W$ the $W$-boson mass.  Thus all top-quark induced FCNC in SM are highly suppressed. Their branching ratios predicted in the SM are of the order of
$10^{-11}$ to $10^{-14}$ \cite{Eilam:1990zc,Mele:1998ag},
far away from present and even future
reaches of either the Large Hadron Collider (LHC) \cite{Beneke:2000hk,
ATLAS} or the International Linear Collider (ILC) \cite{Cobal:2004zt}.

There are many models of NP in which the branching ratios for
the two-body FCNC decays are much larger than those obtained in the SM, and rare top decays might be enhanced to reach detectable levels \cite{reviews}. 
Rare FCNC two-body processes $t \rightarrow c g, \gamma, Z,H$ have been extensively studied in the two-Higgs Doublet Model (2HDM)\cite{Arhrib:2005nx,guasch03, iltan, toscano99}, Alternative Left-Right Symmetric Models \cite{Gaitan:2006eh}
, Constrained \cite{oakes, couture, nanopoulos, petronzio,guasch99} and Unconstrained \cite{misiak,liuli} Minimal Supersymmetric Standard Model (MSSM), Left-Right Supersymmetric Models \cite{Frank:2005vd},  
Supersymmetric Model with R-parity violation \cite{young},  top-assisted technicolor models \cite{wang,lu} as well as models with extra singlet quarks \cite{hou}.

In addition to the two-body rare decays of top quark, some of its
rare three-body decays e.g., $t\to cWW, cZZ, bWZ$ have been
considered in the literature within the SM
\cite{Jenkins:1996zd, Altarelli:2000nt, Bar-Shalom:2005cf}
and for NP \cite{Bar-Shalom:1997sj, Bar-Shalom:2005cf}.
These three body decays are suppressed with respect
to two-body decays in the SM but some of them get
comparably larger within models of NP, such as
two-Higgs-Doublet \cite{Bar-Shalom:1997sj}, especially after
including finite-width effects \cite{Bar-Shalom:2005cf}. In addition, it has been shown that the branching ratio for three-body decay $t \to cgg$  can exceed that of the two body decay $t \to c g$ in both SM   \cite{Eilam:2006uh} and MSSM \cite{Eilam:2006rb}.  The three body decay $t \to cq {\bar q}$ has also been analyzed  \cite{Eilam:2006rb,Cordero-Cid:2004hk}, and while the branching ratio is smaller than that of $t \to cgg$, it is competitive with $t \to cV$.

The above facts strengthens the case for investigating top quark physics further in three  body decays.    
 The goal of the present  paper is to analyze another three-body rare decay, namely
$t\to cl{\bar l}$ in several frameworks and compare it to
both $t\to c \gamma$ and $t\to cq\bar{q},~q=u$. This decay was previously analyzed in \cite{Yue:2004jc} in the framework of topcolor assisted technicolor models, and in \cite{Popov:2006eq} as induced by scalar leptoquarks in a Pati-Salam model.   For b quarks, it is known that under certain circumstances, $b \to s l^+ l^-$ can be more significant than $b \to s \gamma$ in restricting the parameter space, i.e., in supersymmetry in the large $\tan \beta$ region \cite{Huang:2000sm}.

The remainder of
the paper is organized as follows: In Section II we present the calculation
of $t\to cl^+l^-$ in the SM and in Section III we present the same decay in the Two Higgs Doublet Model. Section IV is dedicated to presenting the formalism and calculation of $t\to cl^+l^-$ in the MSSM (in both constrained and unconstrained versions). In each section, we present the Feynman diagrams and analyze the results obtained within the given model. We include, for completeness, the related decay $t \to c \nu_i {\bar \nu}_i$ in Section V. Finally, we analyze the related production of the single top quark in $e^+e^- \to t {\bar c}+{\bar t}c$ at the ILC in Section VI. We conclude and discuss experimental observation in Section VII.

\section{$t \to c l{\bar l}$ in the SM}

\begin{figure}[htb]
        \centerline{ \epsfxsize 6.2in {\epsfbox{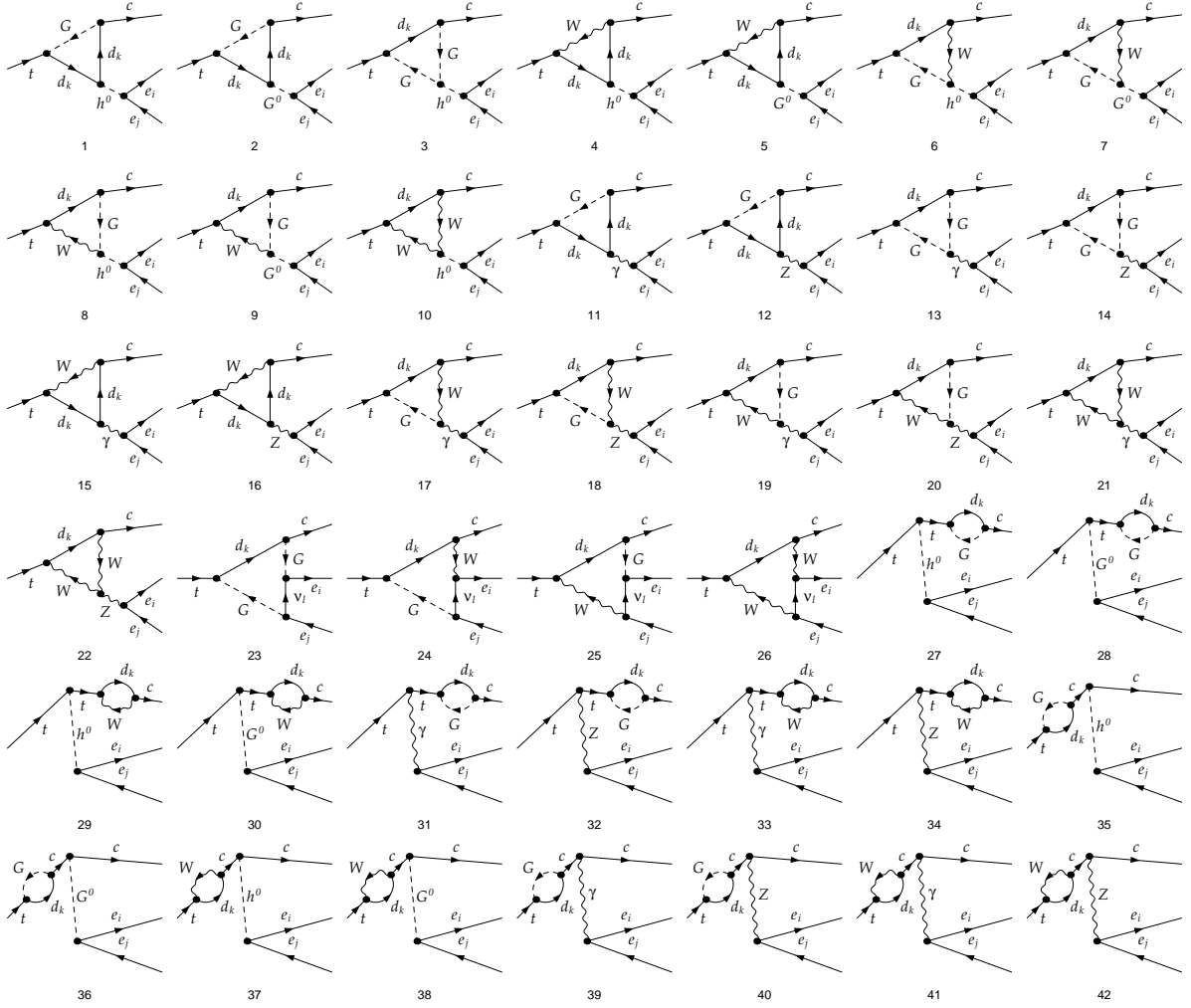}}}
\caption
      {The one-loop SM contributions to $t\to c l^+l^-$ in the
't Hooft-Feynman gauge.}
\label{fig:eeSM}
\end{figure}

In the SM, since FCNC are forbidden, the decay $t \to c l^+l^-$ is induced at one-loop level through charged currents, involving vertices for $Wq q^{\prime}$ in the loop. The Feynman diagrams for this process are depicted in Fig. \ref{fig:eeSM}. 
\begin{figure}[htb]
        \centerline{ \epsfxsize 6.2in {\epsfbox{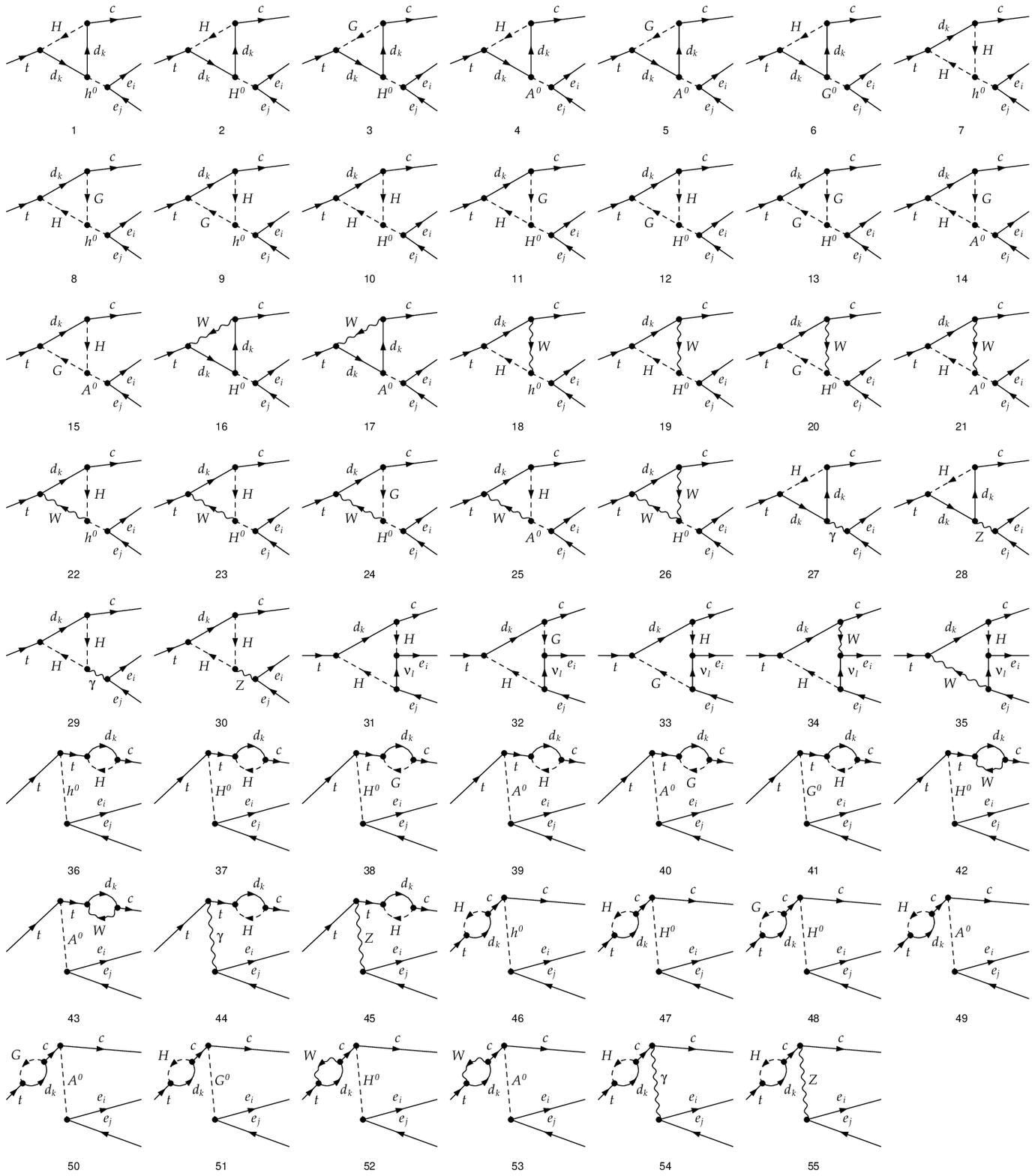}}}
\caption
      {The one-loop 2HDM contributions to $t\to c l^+l^-$ in the
't Hooft-Feynman gauge.}
\label{fig:eeTHDM}
\end{figure}
For our numerical evaluation, we used the following parameters:
\begin{eqnarray}
\!\!\!\!\!\!& & \alpha(m_t)=1/132.5605 \ , \  M_W=80.45\ {\rm GeV}\ ,  \ M_Z=91.1875\ {\rm GeV} \ ,
\ s_W^2=1-M_W^2/M_Z^2\,, \nonumber\\
\!\!\!\!\!\!& & m_t=172.5\ {\rm GeV}, \  m_b=2.85 \ {\rm GeV}, \  m_c=0.63 \ {\rm GeV}, \ \alpha_s(m_t) \approx 0.1068.
\end{eqnarray}
We obtained:
\begin{eqnarray}
BR(t \to c e^+e^-)&=& 8.48 \times 10^{-15},\nonumber \\
BR(t \to c \mu^+ \mu^-)&= & 9.55 \times 10^{-15},\nonumber \\
BR(t \to c \tau^+ \tau^-)&=& 1.91 \times 10^{-14},\nonumber \\
BR(t \to c \sum_i \nu_i {\bar \nu}_i)&=& 2.99 \times 10^{-14}.
\end{eqnarray}
As expected, the branching ratios are very small since even at one loop, the process is suppressed by the Glashow-Iliopoulos-Maiani (GIM) mechanism.\footnote{Though $BR(t \to c e^+e^-)$ is small, it is of same order of magnitude as $BR(t\to cZ)$, and $BR(t\to c\tau^+ \tau^-)$ is of same order of magnitude as $BR(t\to c \gamma)$ \cite{aguilar}.} The SM signal cannot be detected  in current or future high energy collider experiments; thus any signal would be an indication of New Physics.

\section{THE TWO-HIGGS-DOUBLET MODELS}
The Two Higgs Doublet Models (2HDM) are simple extensions of the SM,  formed by adding an extra complex
$SU(2)_L\otimes U(1)_Y$ scalar doublet to the SM Lagrangian.
Motivations for such a extension include the possibility of having CP--violation in the Higgs
sector, and the fact that some models of
dynamical electroweak symmetry breaking
yield the 2HDM as their low-energy effective theory \cite{dewsb}.
There are
three popular versions of this model in the literature, depending on how the two
doublets couple to the fermion sector. Models I
and II (2HDM-I, 2HDM-II) include natural flavor conservation
\cite{2hdm1,hunter}, while model III (2HDM-III) has the simplest
extended Higgs sector that naturally introduces FCNC at the
tree level \cite{sher,antara,reina}.

The most general 2HDM scalar potential which is both 
$SU(2)_L\otimes U(1)_Y$ and CP invariant is given by \cite{hunter}:
\begin{eqnarray}
 V(\Phi_{1}, \Phi_{2})& & =  \lambda_{1} ( |\Phi_{1}|^2-v_{1}^2)^2
+\lambda_{2} (|\Phi_{2}|^2-v_{2}^2)^2+
\lambda_{3}((|\Phi_{1}|^2-v_{1}^2)+(|\Phi_{2}|^2-v_{2}^2))^2 
+\nonumber\\ [0.2cm]
&  & \lambda_{4}(|\Phi_{1}|^2 |\Phi_{2}|^2 - |\Phi_{1}^+\Phi_{2}|^2  )+
\lambda_{5} (\Re(\Phi^+_{1}\Phi_{2})
-v_{1}v_{2})^2+ \lambda_{6} [\Im(\Phi^+_{1}\Phi_{2})]^2\,,
\label{higgspot}
\end{eqnarray}
where $\Phi_1$ and $\Phi_2$ have weak hypercharge Y=1, $v_1$ and
$v_2$ are respectively the vacuum
expectation values of $\Phi_1$ and $\Phi_2$ and the $\lambda_i$
are real--valued parameters. 
Note that this potential violates softly the discrete symmetry
$\Phi_1\to -\Phi_1$, $\Phi_2\to \Phi_2$ by the dimension two term
$\lambda_5 \Re(\Phi^+_{1}\Phi_{2})$.
The above scalar potential has 8 independent parameters
$(\lambda_i)_{i=1,...,6}$, $v_1$ and $v_2$.

After the electroweak symmetry breaks, three of the original
eight degrees of freedom associated to $\Phi_1$ and $\Phi_2$
correspond to the three Goldstone bosons $(G^{\pm}, G^o)$, while
the other five degrees of freedom reduce to five physical Higgs
bosons: $h^0, H^0$ (both CP-even), $A^0$ (CP-odd), and $H^\pm$. Their masses are obtained as usual
by diagonalizing the scalar mass matrix. 
The combination $v_1^2 + v_2^2$ 
is fixed by the electroweak 
scale through $v_1^2 + v_2^2=(2\sqrt{2} G_F)^{-1}$.
 Seven independent parameters are left, which are given in terms of four physical scalar masses $(m_{h^0},
m_{H^0}, m_{A^0}, m_{H^\pm})$, two mixing angles ($\tan \beta = v_1/v_2$
and $\alpha$) and the soft breaking term $\lambda_5$.
 
In this analysis, we impose that, when the independent parameters are varied, the contributions to the $\delta\rho$ parameter from the Higgs
scalars  should not exceed the current limits from precision 
measurements \cite{PDG}: $|\delta\rho|\le 0.001$.

The Feynman rules in the general 2HDM are given in \cite{hunter}. The Feynman diagrams which contribute to $t \to c l^+l^-$ are given in Fig. \ref{fig:eeTHDM}. 
The full one-loop calculation presented here is done in the 't Hooft
gauge with the help of 
 \texttt{FormCalc} \cite{Hahn:2000jm}.
The parameters used are the same as for the standard model. We have tested analyticity for the independent parameters of the model and restricted the parameter space accordingly. The parameter space of the models are severely constrained by the perturbativity constraints ($|\lambda_i|\le 8\pi$). The situation becomes even more severe if one takes into account the unitarity bounds. The branching ratios for $t \to c l^+l^-$ obtained running the 2HDM parameters over their allowed values are all negligibly small (same order as the SM and often numerically indistinguishible), and thus not likely to show up at the present or future colliders. There are few points in the parameter space where $t \to c l^+l^-$ becomes enhanced (around 1-2 orders of magnitude) with respect to the SM predictions but this usually requires very small $\tan\beta$ values ($\le 0.1$) which are excluded \cite{Barger:1989fj}. It is also possible to get some deviation from the SM values if the Higgs bosons ($h^0, H^0, A^0$ and $H^\pm$) are very light (around 100-200 GeV). Overall we do not obtain significant deviations from the SM results and for this reason we do not graph them separately.

\section{The Minimal Supersymmetric Standard Model}

The MSSM constitutes the minimal supersymmetric extension of the
SM. It includes all SM fields, as well as two Higgs doublets needed  to keep anomaly cancellation. One Higgs doublet, $H^1$,
gives mass to the $d$-type fermions (with weak isospin -1/2), the
other doublet, $H^2$, gives mass to the $u$-type fermions (with weak
isospin +1/2). 
All SM multiplets, including the two Higgs doublets, are extended to
supersymmetric multiplets, resulting in scalar partners for quarks and
leptons (squarks and sleptons) and fermionic partners for the
SM gauge bosons and the Higgs bosons (gauginos and higgsinos).
We do not consider effects of complex phases here, i.e.\
we treat all MSSM parameters as real.

The superpotential of the MSSM Lagrangian is
\begin{equation}
     \label{eq:W} {\mathcal{W}} = \mu H^1 H^2 + Y_l^{ij} H^1
{L}^i {e}_R^j + Y_d^{ij} H^1 {Q}^i {d}_R^j
+ Y_u^{ij} H^2 {Q}^i {u}_R^j, 
\label{eq:superpot}
\end{equation}
while the part of the soft-SUSY-breaking Lagrangian responsible for
the non-minimal squark family mixing is given by
\begin{eqnarray}
\label{eq:lagrangian}
\!\!\!\!\!\!\!\mathcal{L}^{\text{squark}}_{\text{soft}} =
-\tilde Q^{i\dagger} (M_{\tilde Q}^2)_{ij} \tilde Q^j
-\tilde u^{i\dagger} (M_{\tilde U}^2)_{ij} \tilde u^j
-\tilde d^{i\dagger} (M_{\tilde D}^2)_{ij} \tilde d^j 
+ Y_u^i A_u^{ij} \tilde Q_i H^2 \tilde u_j
+ Y_d^i A_d^{ij} \tilde Q_i H^1 \tilde d_j.
\end{eqnarray}
In the above expressions $ Q$ is the $SU(2)$ scalar doublet, $ u$, $ d$ are
the up- and down-quark $SU(2)$ singlets ($\tilde Q, \tilde u, \tilde d$
represent scalar quarks), respectively, $Y_{u,d}$ are the
Yukawa couplings and $i,j$ are generation indices. Here $A^{ij}$ are the 
trilinear scalar couplings and $H^{1,2}$ represent two $SU(2)$
Higgs doublets with vacuum expectation values
\begin{equation}
     \langle H^1 \rangle = \left( \begin{array}{c}
\frac{v_1}{\sqrt{2}} \\ 0 \end{array} \right) \equiv \left(
\begin{array}{c} \frac{v \cos \beta}{\sqrt{2}} \\ 0 \end{array}
\right), \hspace{0.8cm} \langle H^2 \rangle = \left(
\begin{array}{c} 0 \\ \frac{v_2}{\sqrt{2}} \end{array} \right) \equiv
\left( \begin{array}{c} 0 \\ \frac{v \sin \beta}{\sqrt{2}}
\end{array} \right),
\end{equation}
where $v=(\sqrt{2}G_F)^{-1/2}=246$~GeV, and the angle $\beta$ is
defined by $\tan \beta\equiv v_2/v_1$, the ratio of the vacuum
expectation values of the two Higgs doublets; and $\mu$ is the Higgs
mixing parameter.

The squark mass term of the MSSM Lagrangian is given by
\begin{equation}
{\cal L}_{mass}^f = -\frac{1}{2} 
   \Big( \tilde{f}_L^{\dag},\tilde{f}_R^{\dag} \Big)\; {\cal M}_{\tilde f}^2 \; 
   \left( \begin{array}{c} \tilde{f}_L \\[0.5ex] \tilde{f}_R \end{array}\right) ~,
\label{squarkmassmatrix}
\end{equation}
where we assume the following form of ${\cal M}_{\tilde f}^2$
\begin{equation}
\label{eq:squarkmass}
\!\!\!\!\!\!\!\!{\cal M}^2_{\tilde {u}\{\tilde d\} }=
\left( \begin{array}{cccccc}
M_{{\tilde L} u\{d\}}^2 & 0 & 0 & m_{u\{d\}} {\cal A}_{u\{d\}} & 0 & 0  \\
0 & M_{{\tilde L} c\{s\}}^2 & (M^2_{\tilde U\{\tilde D\}})_{LL} & 0 & m_{c\{s\}} {\cal A}_{c\{s\}}
&(M^2_{\tilde U\{\tilde D\}})_{LR} \\
0 & (M^2_{\tilde U\{\tilde D\}})_{LL} & M_{{\tilde L} t\{b\}}^2 & 0 &
(M^2_{\tilde U\{\tilde D\}})_{RL} & m_{t\{b\}} {\cal A}_{t\{b\}} \\[.3ex]
m_{u\{d\}} {\cal A}_{u,\{d\}}& 0 & 0 & M_{{\tilde R} u\{d\}}^2 & 0 & 0 \\
0 & m_{c\{s\}} {\cal A}_{c\{s\}} & (M^2_{\tilde U\{\tilde D\}})_{RL} & 0 &M_{{\tilde R} c\{s\}}^2 &
(M^2_{\tilde U\{\tilde D\}})_{RR} \\
0 & (M^2_{\tilde U\{\tilde D\}})_{LR} & m_{t \{b\}} {\cal A}_{t \{b\}} & 0 &
(M^2_{\tilde U \{\tilde D\}})_{RR} &M_{{\tilde R} t \{b\}}^2
\end{array} \right)
\end{equation}
with
\begin{eqnarray}
\label{eq:squarkparam}
M_{{\tilde L}q}^2 &=&
      M_{\tilde Q,q}^2 + m_q^2 + \cos2\beta (T_q - Q_q s_W^2) M_Z^2\,,
\nonumber \\
M_{{\tilde R}\{u,c,t\}}^2 &=&
      M_{\tilde U,\{u,c,t\}}^2 + m_{u,c,t}^2 + \cos2\beta Q_t s_W^2
M_Z^2\,, \nonumber \\
M_{{\tilde R}\{d,s,b\}}^2 &=&
      M_{\tilde D,\{d,s,b\}}^2 + m_{d,s,b}^2 + \cos2\beta Q_b s_W^2 M_Z^2\,, \\
{\cal A}_{u,c,t} &=& A_{u,c,t} - \mu\cot\beta\,,\;\;\;\;
{\cal A}_{d,s,b} = A_{d,s,b} - \mu\tan\beta\,, \nonumber
\end{eqnarray}
with $m_q$, $T_q$, $Q_q$ the mass, isospin, and electric charge of the
quark $q$, $M_Z$ the $Z$-boson mass, $s_W \equiv \sin\theta_W$ and
$\theta_W$ the electroweak mixing angle. Since we are concerned with 
top quark physics, we assume that the
 squark mixing is significant only for
transitions between the
squarks of the second and third generations.  These mixings are expected to be
the largest in Grand Unified Models  and are also experimentally the
least constrained.  The most stringent bounds on these transitions
come from $b \to s \gamma $.  In
contrast, there exist strong experimental bounds involving the first
squark generation, based on data from $K^0$--$\bar K^0$ and $D^0$--$\bar
D^0$ mixing~\cite{Gabbiani:1996hi}.

We define the dimensionless flavor-changing parameters
$(\delta_{U,D}^{23})_{AB}$ $(A,B = L,R)$ from the
flavor off-diagonal elements of the squark mass matrices in the following
way. First, to simplify the calculation we assume
that all diagonal entries $M^2_{\tilde{Q},q}$ and $M^2_{\tilde{U}(\tilde{D}),q}$
 are set equal to the common value
$M^2_{\rm{SUSY}}$, and then we normalize the off-diagonal elements
to $M^2_{\rm{SUSY}}$  \cite{Harnik:2002vs,Besmer:2001cj},
\begin{eqnarray}
(\delta_{U(D)}^{ij})_{AB} =
\frac{(M^2_{\tilde{U}(\tilde{D})})_{AB}^{ij}}{M^2_{\rm{SUSY}}},\;\; (i\ne j,\;\; i,j=2,3\;\;\;A,B=L,R).
 \label{deltadefb}
\end{eqnarray}
     The matrix ${\cal M}^2_{\tilde{u}}$ in Eq.~(\ref{eq:squarkmass}) can
further be diagonalized by an additional $6\times 6$ unitary
matrix $\Gamma_U$ to give the up squark mass eigenvalues
\begin{eqnarray}
\left({\cal M}^2_{\tilde{u}}\right)^{\text{diag}} = \Gamma_U^{\dagger}
{\cal M}^2_{\tilde{u}} \Gamma_U
     \label{eq:gammaudef}.
\end{eqnarray}
For the down squark mass matrix, we also can define
${\mathcal{M}}_{\tilde{d}}^2$ as the similar form of
Eq.~(\ref{eq:gammaudef}) with the replacement of
$(M^2_{\tilde{U}})_{AB}$ ($A,B=L,R$) by $(M^2_{\tilde{D}})_{AB}$ and $\Gamma_U$ by $\Gamma_D$.
Note that $SU(2)_L$ gauge invariance implies that
$(M^2_{\tilde{U}})_{LL} = K_{CKM} (M^2_{\tilde{D}})_{LL} K_{CKM}^\dagger$ and the
matrices $(M^2_{\tilde{U}})_{LL}$ and $(M^2_{\tilde{D}})_{LL}$ are
correlated. Since the connecting equations are rather complicated and
contain several unknown parameters, we proceed by including the
flavor changing parameters $(\delta_{U(D)}^{ij})_{AB} $ as independent
quantities, while restricting them using previously set bounds
\cite{Gabbiani:1996hi}. Unlike some other approaches \cite{Hall:1985dx},
the mass matrix in
Eq.~(\ref{eq:gammaudef}) (and the similar one in the down-sector) is
diagonalized and the flavor changing parameters enter into our
expressions through the matrix $\Gamma_{U,D}$. So, in the top
decays $t\to cl^+l^-$, the new flavor changing neutral currents show
themselves in both gluino-squark-quark and neutralino-squark-quark
couplings in the up-type squark loops and in the
chargino-squark-quark coupling in the down-type squark loops.
 We briefly review the ingredients of the flavor-violation in each sector, 
then give the interactions responsible for the violation.

\begin{figure}[htb]
        \centerline{\epsfxsize 6.5in {\epsfbox{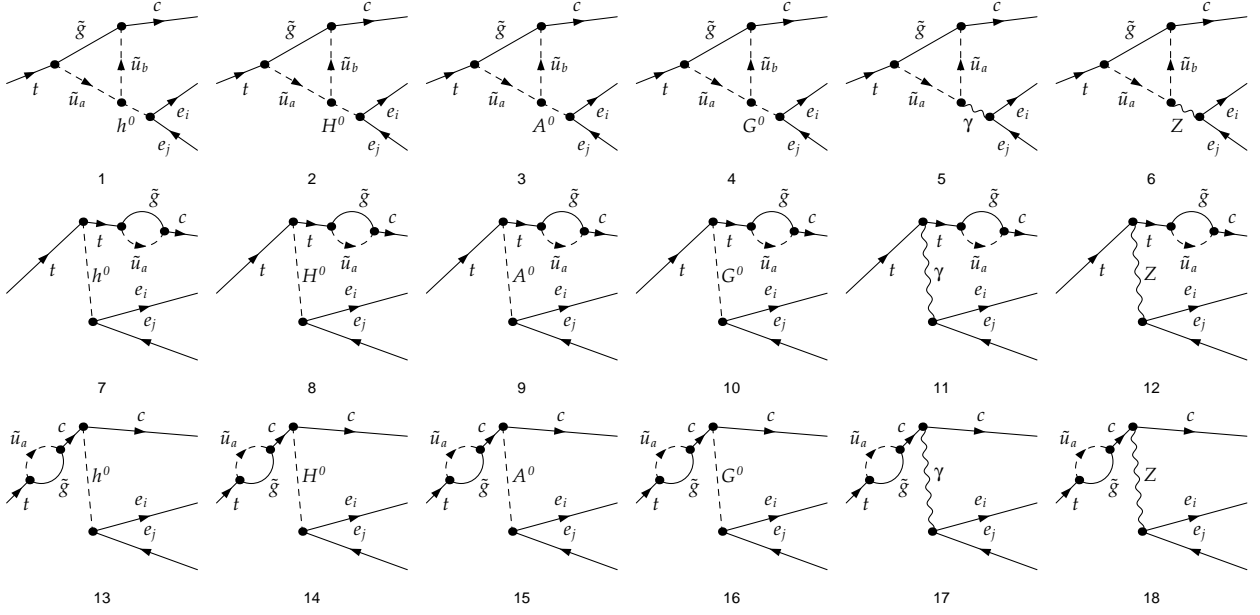}}}
\caption
      {The one-loop gluino contributions to $t\to c l^+l^-$ in MSSM in the
't Hooft-Feynman gauge.}
\label{fig:eegluino}
\end{figure}

The gluino, $\tilde g$, is the spin~1/2 superpartner (Majorana fermion) of
the gluon. According to the generators of $SU(3)_C$ (color octet),
there are 8 gluinos, all having the same Majorana mass
\begin{equation}
m_{\tilde g} = M_3~. 
\end{equation}
In the super-CKM basis, the quark-up squark-gluino ($\tilde g$)
interaction is given by
\begin{equation}
\mathcal{L}_{u \tilde{u} \tilde g}= \sum_{i=1}^{3}\sqrt{2}\, g_s \,
T^r_{st} \left[ \bar
u^{s}_i \,(\Gamma_U)^{ia}\,P_L\, \tilde g^r \,\tilde u^{t}_a - \bar
u^{s}_i \,(\Gamma_U)^{(i+3)a}\,P_R \,\tilde g^r \,\tilde u^{t}_a +
\text{H.c.} \right]\,,
\end{equation}
where $T^{r}$ are the $SU(3)_{c}$ generators,  $P_{L,R}\equiv (1\mp
\gamma_5)/2$, $i=1,2,3$ is the generation index, $a=1, \ldots, 6$ is
the scalar quark index, and $s,t$ are color indices. In the gluino
interaction, the flavor changing effects from soft broken
terms $M^2_{\tilde Q}$, $M^2_{\tilde U}$ and $A_{u}$ on the observables are
introduced through the matrix $\Gamma_U$. The Feynman graphs generated by gluino contributions to the decay $t\to c l^+l^-$ are depicted in Fig \ref{fig:eegluino}.

The charginos $\tilde{\chi}_i^+\; (i=1,2)$ are four component Dirac
fermions. The mass eigenstates are obtained from the winos
$\tilde{W}^\pm$ and the charged higgsinos $\tilde{H}^-_1$,
$\tilde{H}^+_2$:
\begin{equation}
\tilde{W}^+ = \left( \begin{array}{c} -i \lambda^+ \\[0.5ex] i \bar{\lambda}^- \end{array}\right) 
             ;\quad
\tilde{W}^- = \left( \begin{array}{c} +i \lambda^- \\[0.5ex] i \bar{\lambda}^- \end{array}\right) 
            ;\quad
\tilde{H}^+_2 = \left( \begin{array}{c} \psi^+_{H_2} \\[0.5ex] \bar{\psi}^-_{H_1} \end{array}\right) 
               ;\quad
\tilde{H}^-_1 = \left( \begin{array}{c} \psi^-_{H_1} \\[0.5ex] \bar{\psi}^+_{H_2} \end{array}\right) ~.
\end{equation}
The chargino masses are defined as mass eigenvalues of the
diagonalized mass matrix,
\begin{equation}
{\cal L}^{\tilde{\chi}^+}_{\rm mass} = -\frac{1}{2}\,
         \Big( \psi^+,\psi^- \Big) \left( \begin{array}{cc} 0 & {\bf X}^T \\ {\bf X} & 0  \end{array}\right)\left( \begin{array}{c} \psi^+ \\ \psi^-  \end{array}\right)+{\rm H.c.}
\end{equation}
or given in terms of two-component fields
\begin{equation}
\left. \begin{array}{c} 
       \psi^+ = (-i\lambda^+, \psi^+_{H_2}),\;\;\;\;
       \psi^- = (-i\lambda^-, \psi^-_{H_1}) 
       \end{array} 
\right.~,
\end{equation}
where {\bf X} is given by
\begin{equation}
{\bf X} = \left( \begin{array}{cc} M_2 & \sqrt2\, M_W \sin \beta \\[1ex] \sqrt2\, M_W\, 
          \cos \beta & \mu  \end{array}\right)~.
\end{equation}
In the mass matrix, $M_2$ is the soft SUSY-breaking parameter for the
Majorana mass term. and $\mu$ is the Higgsino mass parameter from the Higgs
potential. 

The physical (two-component) mass eigenstates are obtained via
unitary $(2 \times 2)$~matrices $U$ and $V$:
\begin{equation}
\left. \begin{array}{c} 
       \chi_i^+ = V_{ij}\, \psi_j^+,\;\;\;\;
       \chi_i^- = U_{ij}\, \psi_j^- 
       \end{array} 
\right. \qquad i,j=1,2~.
\end{equation}
The eigenvalues
of the diagonalized matrix
\begin{eqnarray}
{\bf M}_{{\rm diag}}^{\tilde{\chi}^+} &=& 
{U^*\, {\bf X}\, V}^{-1} \; =\; 
 \left (\begin{array}{cc}m_{\tilde{\chi}_1^+} & 0 \\[0.5ex] 0 & m_{\tilde{\chi}_2^+} \end{array} \right)\,,
\end{eqnarray}
are given by
\begin{eqnarray}
m^2_{\tilde{\chi}_{1,2}^+} &=& \frac{1}{2}\, \bigg\{ 
    M_2^2 + \mu^2 + 2M_W^2 \mp \Big[ (M_2^2-\mu^2)^2\nonumber \\ 
& &  +\; 4M_W^4\cos^2 2\beta + 4M_W^2(M_2^2+\mu^2+2\,\mu\, M_2\, \sin 2 \beta) 
    \Big]^{\frac{1}{2}} \bigg\}~.
\label{Charmasse}
\end{eqnarray}
The relevant Lagrangian terms for the
     quark-down squark-chargino ($\tilde {\chi}^{\pm}_\sigma$)
interaction are given by
\begin{eqnarray}
\mathcal{L}_{u\tilde{d}\tilde{\chi}^{+}}\!\!\!&=\!\!\!&\sum_{\sigma=1}^{2}\,
\sum_{i,j=1}^{3}\left\{ \bar{u}
_ {i}\,[V_{\sigma 2}^{*}\,(Y_{u}^{\text{diag}}\,K_{CKM})_{ij}]
\,P_L\,\tilde{\chi}
_{\sigma}^{+}\,(\Gamma_D)^{ja}\,\tilde{d}_{a}-\bar{u}_{i}\,[g\,U_{\sigma
1}\,(K_{CKM})_{ij}]\, P_R\,
\tilde{\chi}_{\sigma}^{+}\,(\Gamma_D)^{ja} \,\tilde{d}_a\right.   \nonumber \\
& &  \left. +\,\bar{u}_{i}\,[U_{\sigma 2}\,(K_{CKM}\,Y_{d}^{%
\text{diag}})_{ij}]
\,P_R\,\tilde{\chi}_{\sigma}^{+}\,(\Gamma_D)^{(j+3)a}\,\tilde{d}_a
\right\} +\text{%
H.c.}\,,
\end{eqnarray}
\begin{figure}[htb]
        \centerline{\epsfxsize 6.5in {\epsfbox{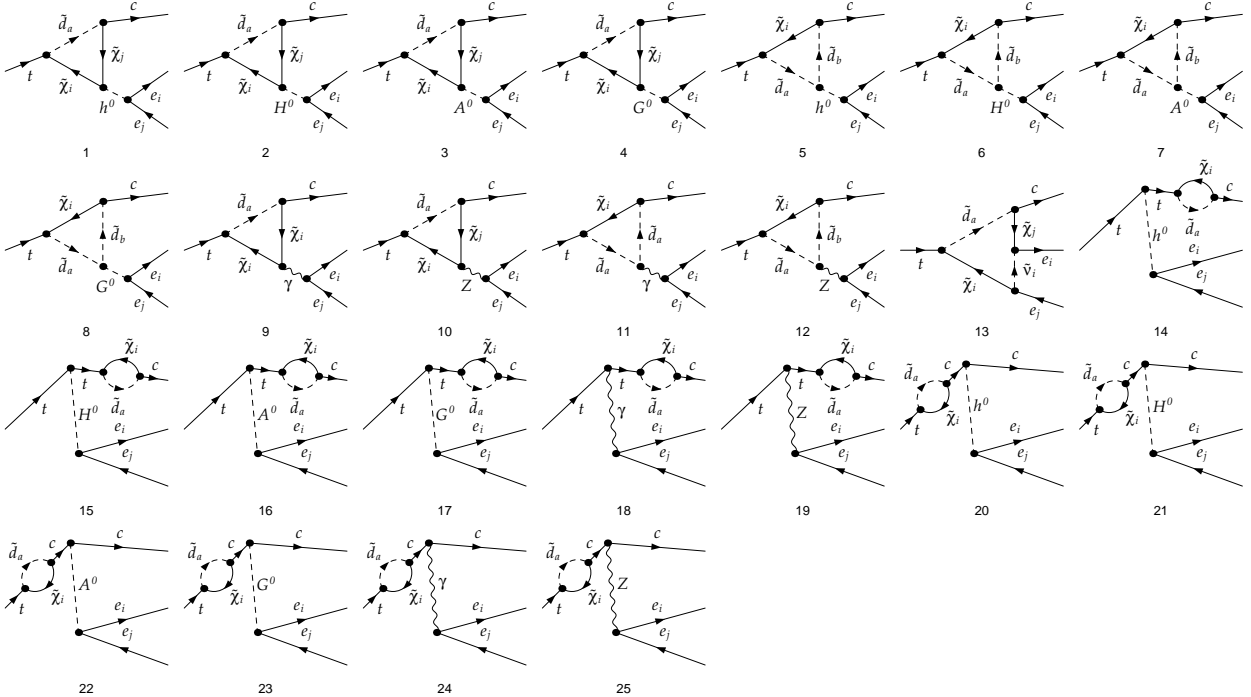}}}
\caption
      {The one-loop chargino contributions to $t\to c l^+l^-$ in MSSM in the
't Hooft-Feynman gauge.}
\label{fig:eechargino}
\end{figure}
where the index $\sigma$ refers to chargino mass eigenstates.
$Y_{u,d}^{\text{diag}}$ are the
diagonal up- and down-quark Yukawa couplings, and $V$, $U$ the 
chargino rotation matrices defined by $U^{*}M_{\tilde {\chi} 
^{+}}V^{-1}=\mathrm{diag}%
(m_{\tilde {\chi} _{1}^{+}},m_{\tilde {\chi} _{2}^{+}})$. The flavor
changing effects
arise from both the off-diagonal elements in the CKM matrix $K_{CKM}$
and from the soft supersymmetry breaking terms in $\Gamma_D$. The Feynman graphs generated by chargino contributions to the decay $t\to c l^+l^-$ are shown in Fig \ref{fig:eechargino}.

\begin{figure}[htb]
        \centerline{\epsfxsize 6.5in {\epsfbox{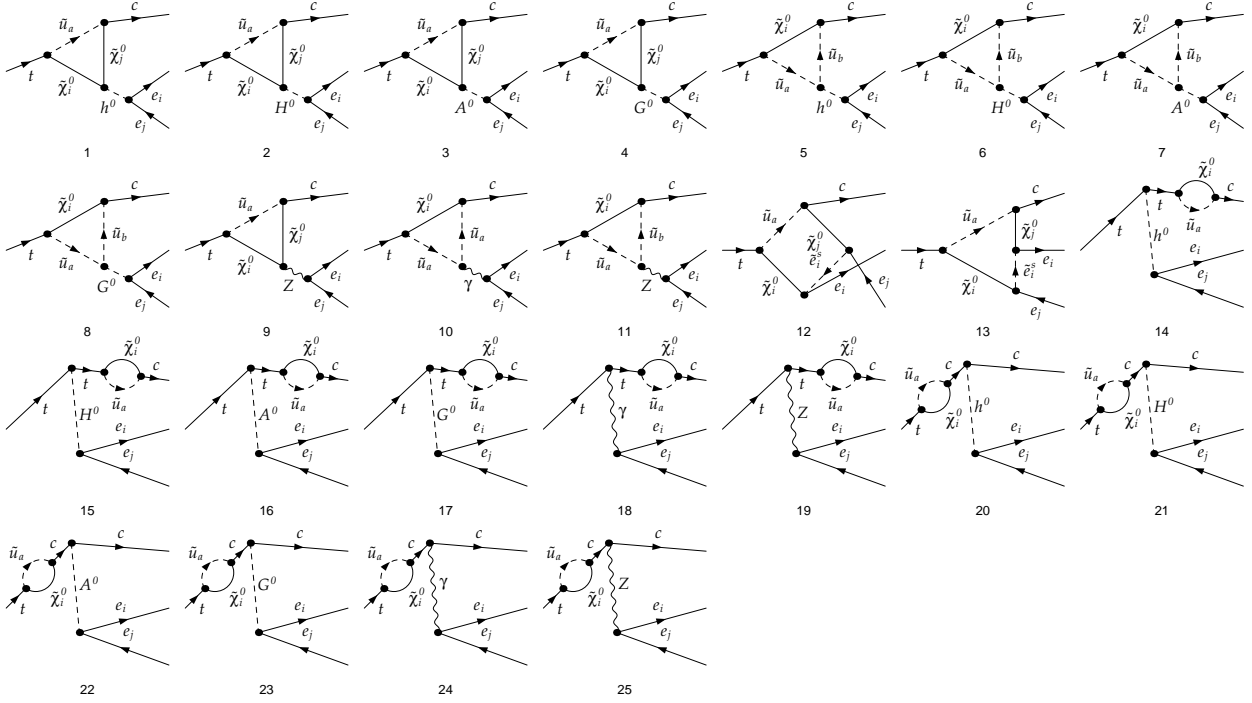}}}
\caption
      {The one-loop neutralino contributions to $t\to c l^+l^-$ in MSSM in the
't Hooft-Feynman gauge.}
\label{fig:eeneutralino}
\end{figure}

Neutralinos $\tilde{\chi}_i^0\; (i=1,2,3,4)$ are four-component
Majorana fermions. They are the mass eigenstates of the
photino,~$\tilde{\gamma}$, the zino,~$\tilde Z$, and the neutral higgsinos,
$\tilde{H}^0_1$ and $\tilde{H}^0_2$, with
\begin{equation}
\tilde{\gamma} = \left( \begin{array}{c} -i \lambda_\gamma \\[0.5ex]
                      i \bar{\lambda}_\gamma \end{array}\right);\quad
\tilde{Z} =\left( \begin{array}{c} -i \lambda_Z \\[0.5ex] i \bar{\lambda}_Z \end{array}\right)
            ;\quad
\tilde{H}^0_1 = \left( \begin{array}{c} \psi^0_{H_1} \\[0.5ex] \bar{\psi}^0_{H_1} \end{array}\right)
              ;\quad
\tilde{H}^0_2 = \left( \begin{array}{c} \psi^0_{H_2} \\[0.5ex] \bar{\psi}^0_{H_2} \end{array}\right)~.
\end{equation}
The mass term in the Lagrangian density is given by
\begin{equation}
{\cal L}_{\tilde{\chi}^0,{\rm mass}} = -\frac{1}{2}(\psi^0)^T\, {\bf Y}\,
                                  \psi^0 + {\rm H.c.}
\end{equation}
with the two-component fermion fields
\begin{equation}
(\psi^0)^T = (-i\lambda^\prime , -i\lambda^3 , \psi_{H_1}^0 ,
              \psi_{H_2}^0)~.
\end{equation}
The mass matrix {\bf Y} is given by
\begin{equation}
\renewcommand{\arraystretch}{1.2}
\!\!\!\!{\bf Y} = \left( \begin{array}{cccc} M_1 & 0 & -M_Z\sin \theta_W\cos \beta & M_Z\sin \theta_W\sin \beta \\ 0 &
          M_2 & M_Z\cos \theta_W\cos \beta & -M_Z\cos \theta_W\sin \beta \\
          -M_Z\sin \theta_W\cos \beta & M_Z\cos \theta_W\cos \beta & 0 & -\mu \\ M_Z\sin \theta_W\sin \beta &
          -M_Z\cos \theta_W\sin \beta& -\mu & 0\end{array}\right)\,.
\label{Y}
\end{equation}
The physical neutralino mass eigenstates are obtained with the
unitary transformation matrix~$N$:
\begin{equation}
\chi_i^0 = N_{ij}\, \psi_j^0 \qquad i,j=1,\ldots,4,
\end{equation}
The diagonal mass matrix is then given by
\begin{equation}
{\bf M}_{{\rm diag}}^{\tilde{\chi}^0} = {N^*\,{\bf Y}\, N}^{-1}~.
\end{equation}
the relevant Lagrangian terms for the quark-up squark
neutralino ($\tilde {\chi}^{0}_n$) interaction are
\begin{eqnarray}
\mathcal{L}_{u\tilde{u}\tilde{\chi}^{0}}&=&\sum_{n=1}^{4}\sum_{i=1}^{3}\left\{
\bar{u}
_{i}\,N_{n1}^{*}\,\frac{4}{3}\frac{g}{\sqrt{2}}\tan \theta _{W} \,P_L\,\tilde{%
\chi}_{n}^{0}\,(\Gamma_U)^{(i+3)a}\,\tilde{u}_a-\bar{u}_{i}\,N_{n4}^{*}\,Y_{u}^{\text{%
diag}}\,P_L\,\tilde{\chi}_{n}^{0}\,(\Gamma_U)^{ia}\,\tilde{u}_a
\right.   \nonumber \\
&-& \left.\bar{u}_{i}\,\frac{g}{\sqrt{2}}\left( N_{n2}+%
\frac{1}{3}N_{n1}\tan \theta _{W}\right)\,P_R
\,\tilde{\chi}_{n}^{0}\,(\Gamma_U)^{ia}\,\tilde{u}_a%
-\bar{u}_{i}\,N_{n4}\,Y_{u}^{\text{diag}}\,P_R\,\tilde{\chi}_{n}^{0}\,(\Gamma_U)^{(i+3)a}\,%
\tilde{u}_a\right\} \,,\nonumber \\
\end{eqnarray}
where $N$ is the rotation matrix which diagonalizes the
neutralino mass matrix $M_{\tilde \chi^0}$, $N^{*}M_{\tilde
\chi^0}N^{-1}=\mathrm{diag}(m_{%
\tilde {\chi}_{1}^{0}},\,m_{\tilde {\chi}_2^0}, \,m_{\tilde
{\chi}_3^0}, \,m_{\tilde {\chi}_4^0})$. As in
the gluino case, FCNC terms arise only from supersymmetric parameters
in $\Gamma_U$. The Feynman graphs generated by neutralino contributions to the decay $t\to c l^+l^-$ are given  in Fig \ref{fig:eeneutralino}.

Note that MSSM can be studied  as a model in itself, or as a low energy realization of a supersymmetric grand unified scenario (SUSY GUT). In SUSY GUTs $M_1$, $M_2$ and $M_3$ are not independent but connected
via
\begin{equation}
m_{\tilde g} = M_3 = \frac{g_3^2}{g_2^2}\, M_2 \; = \;
      \frac{\alpha_s}{\alpha_{\rm em}}\, \sin \theta_W^2\, M_2, \;\;
M_1 = \frac{5}{3} \frac{\sin \theta_W^2}{\cos \theta_W^2}\, M_2~.
\label{G-GUT}
\end{equation}
which results in a reduction of the number of independent parameters. We shall refer to this as the mSUGRA scenario and label the figures accordingly.

In the unconstrained MSSM no specific assumptions are made about the
underlying SUSY-breaking mechanism, and a parametrization of all
possible soft SUSY-breaking terms is used that does not alter the relation
between the dimensionless couplings (which ensures that the absence of
quadratic divergences is maintained). This parametrization has the 
advantage of being very general, but the disadvantage of introducing more
than 100 new parameters in addition to the SM. While in principle these
parameters (masses, mixing angles, complex phases) could be chosen
independently of each other, experimental constraints from
flavour-changing neutral currents, electric dipole moments, etc.\
seem to favour a certain degree of universality among the soft
SUSY-breaking parameters. 
\begin{figure}[htb]
\begin{center}$
	\begin{array}{cc}
\hspace*{-0.5cm}
	\includegraphics[width=3.1in]{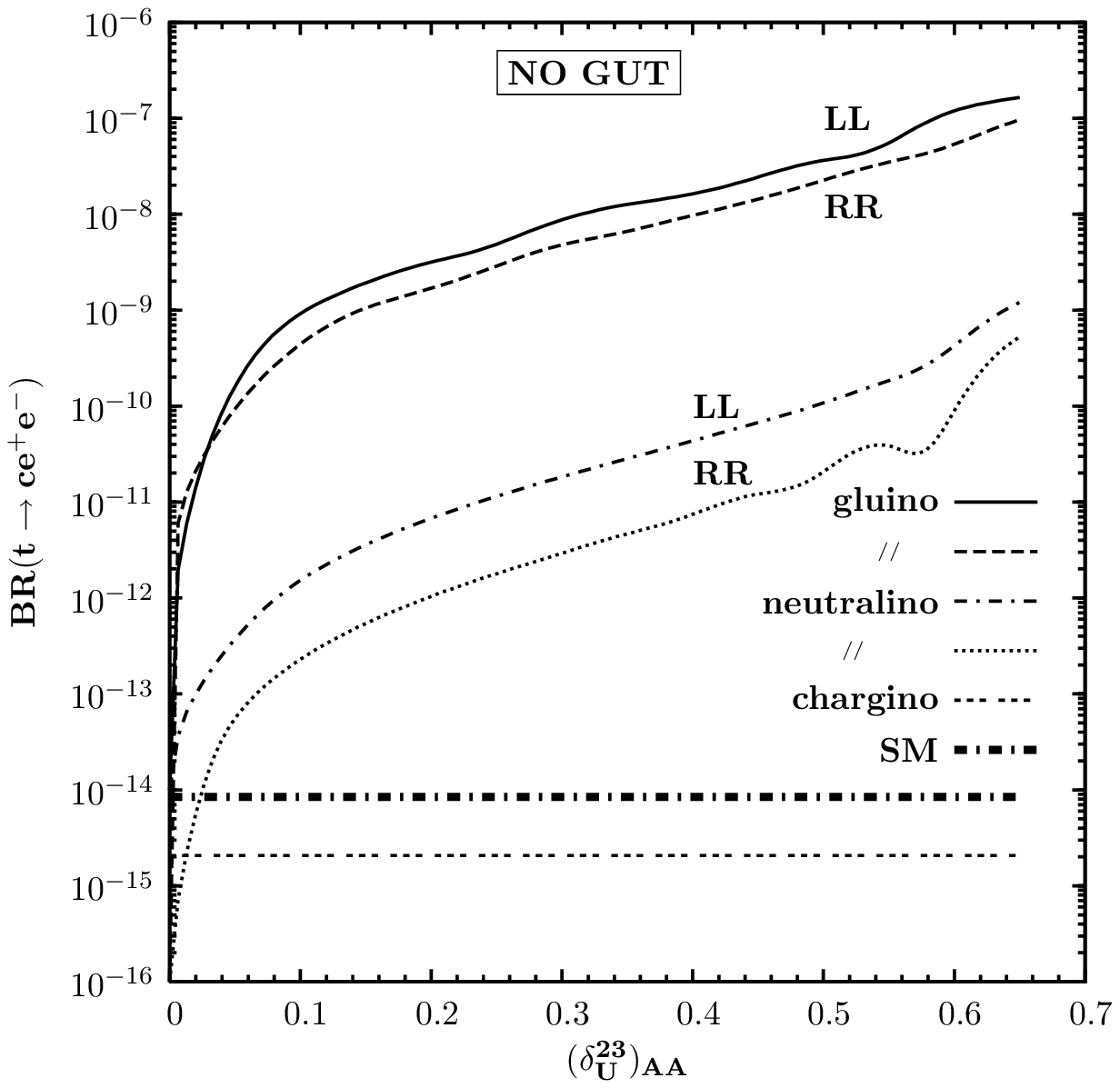} &
	\includegraphics[width=3.1in]{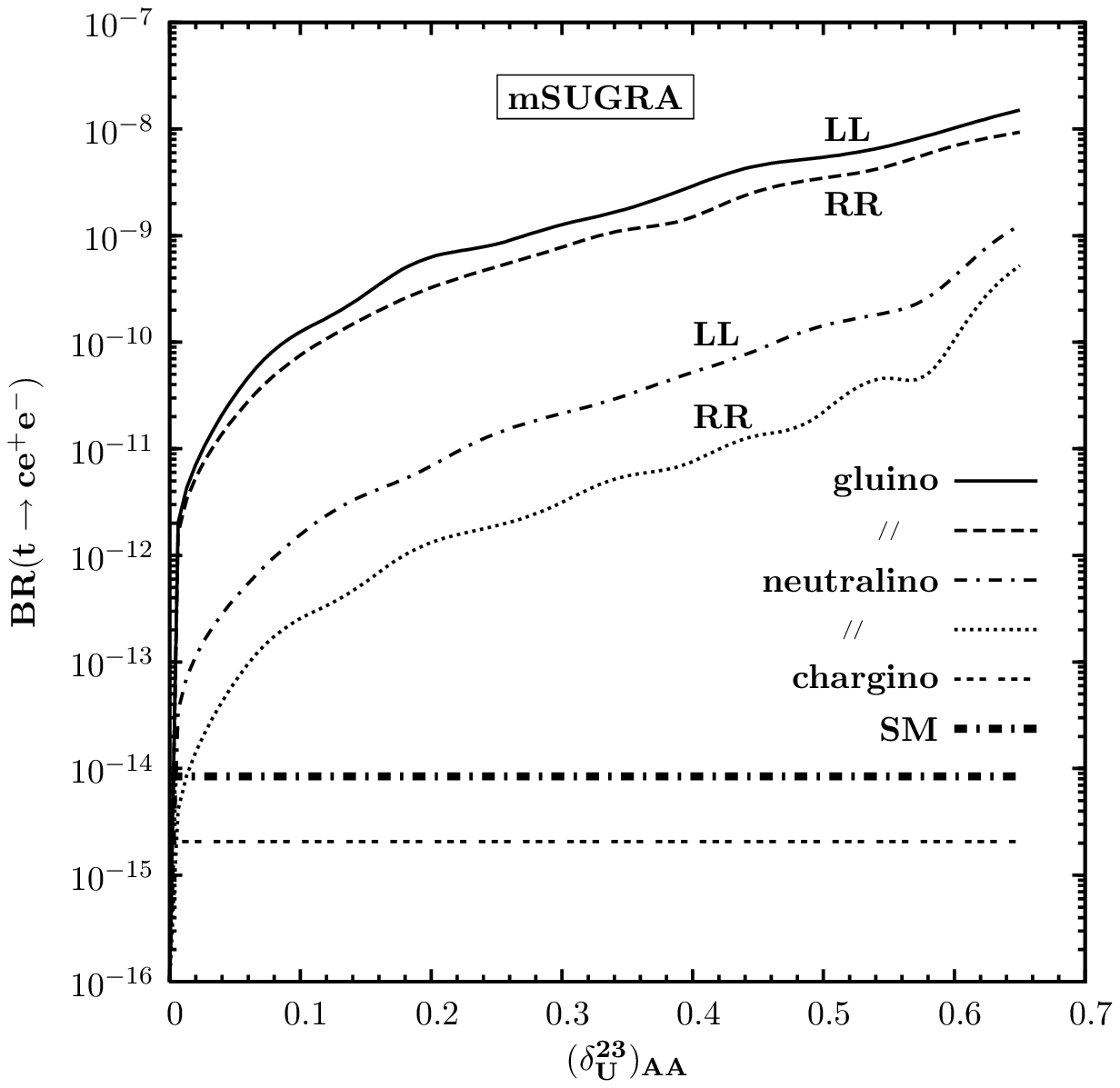} \\
\hspace*{-0.5cm}
	\includegraphics[width=3.1in]{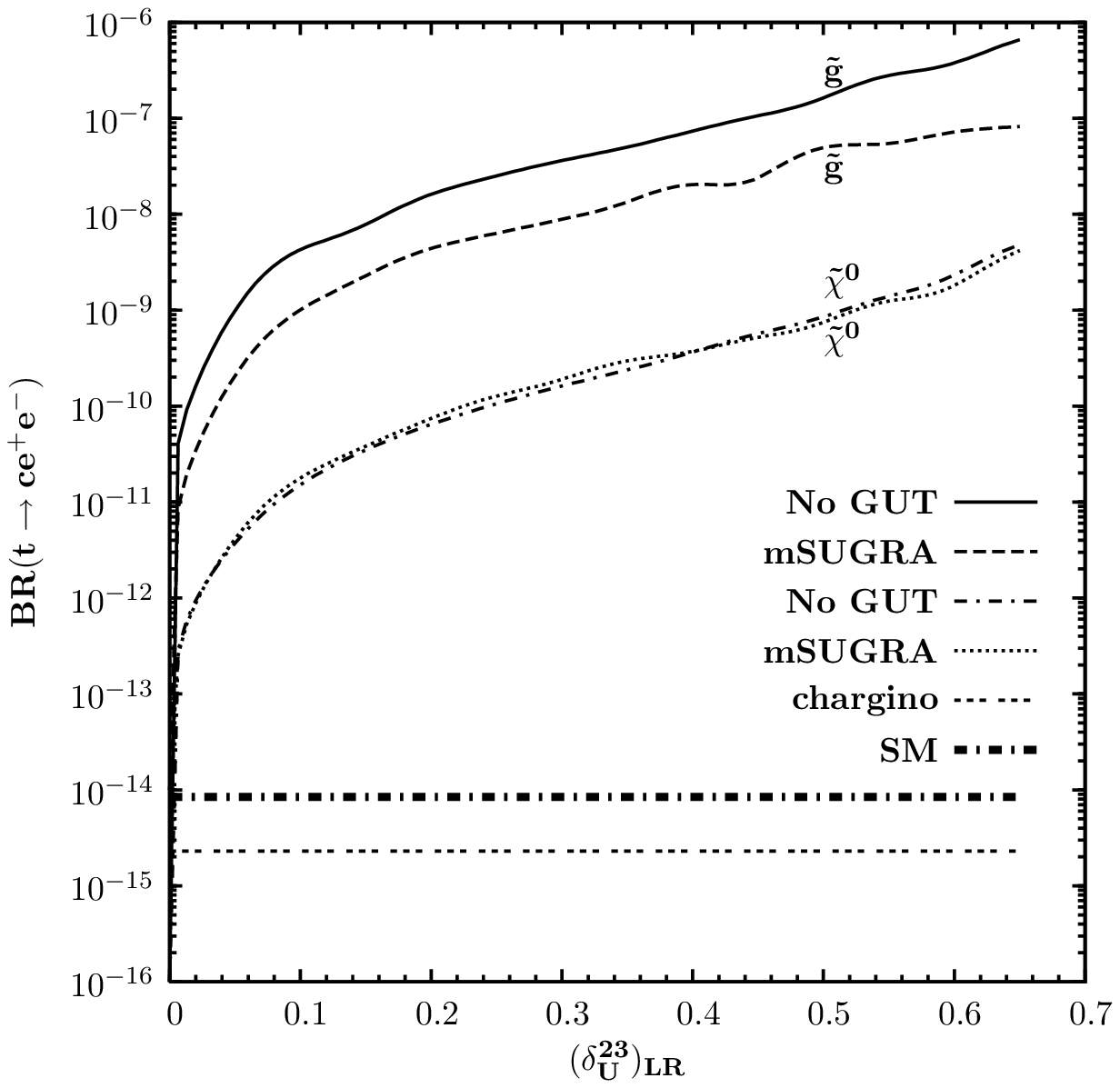} &
	\includegraphics[width=3.1in]{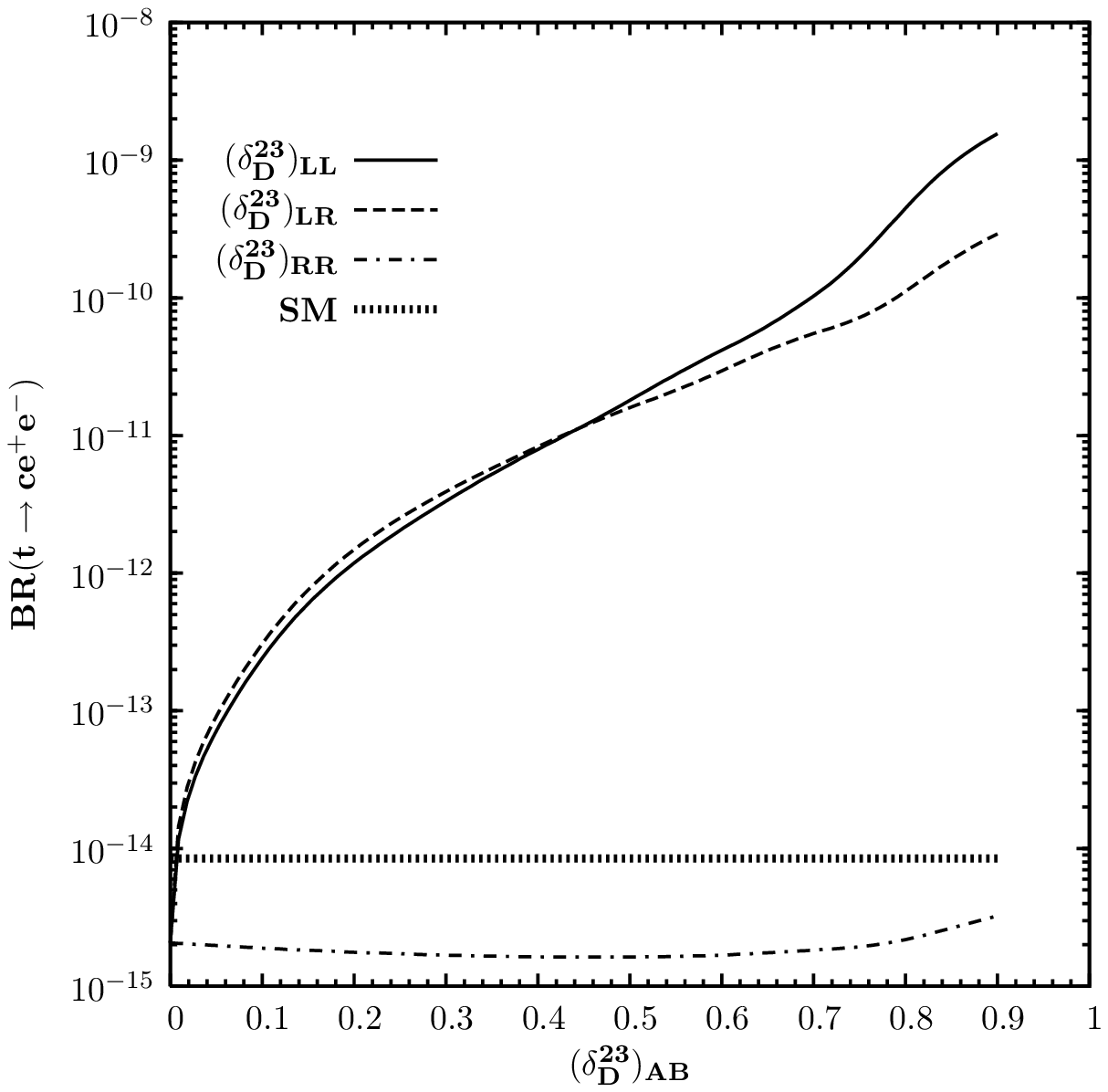}
	\end{array}$
\end{center}
\vskip -0.2in
     \caption{
\underline{Upper left panel}:
BR's of $t\to ce^+e^-$ decay as
function of $(\delta^{23}_U)_{AA}$ without the GUT
relations ($m_{\tilde{g}}=250$ GeV).
\underline{Upper right panel}:
BR's as function of $(\delta^{23}_U)_{AA}$
under the same conditions in mSUGRA scenarios.
\underline{Lower left panel}:
BR's of $t\to ce^+e^-$ decay as
function of $(\delta^{23}_U)_{LR}$.
\underline{Lower right panel}:
BR's as function of $(\delta^{23}_D)_{AB},\,A,B=L,R$.
The parameters are chosen as
$\tan\beta=10$, $m_{A^0}=M_{\rm {SUSY}}=500$
GeV, $M_2=\mu=200$ GeV, and $A_t=1.2$ TeV.}\label{fig:tceeFV}
\end{figure}
\begin{figure}[htb]
\begin{center}$
    \begin{array}{cc}
\hspace*{-0.5cm}
    \includegraphics[width=3.1in]{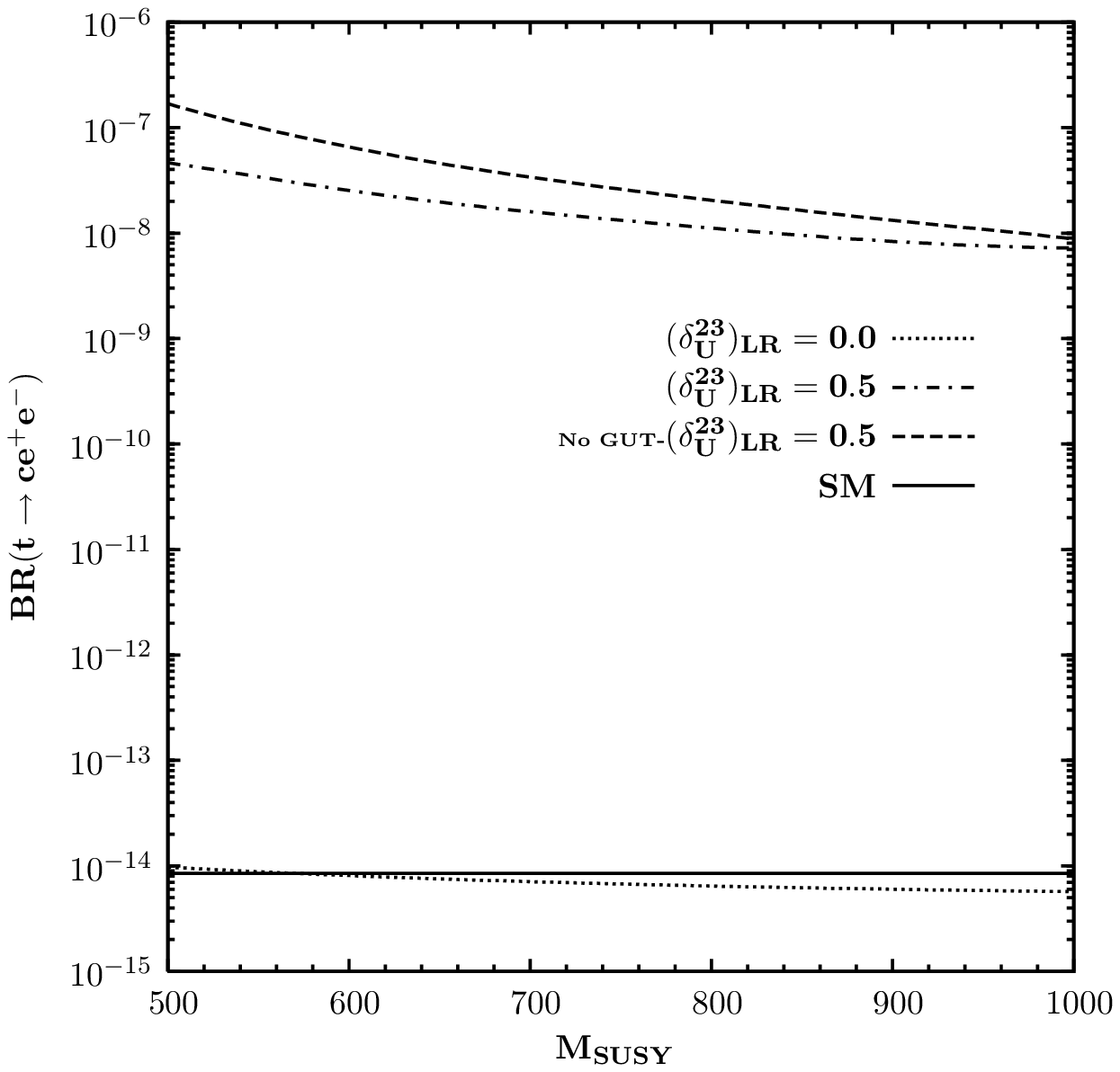} &
    \includegraphics[width=3.1in]{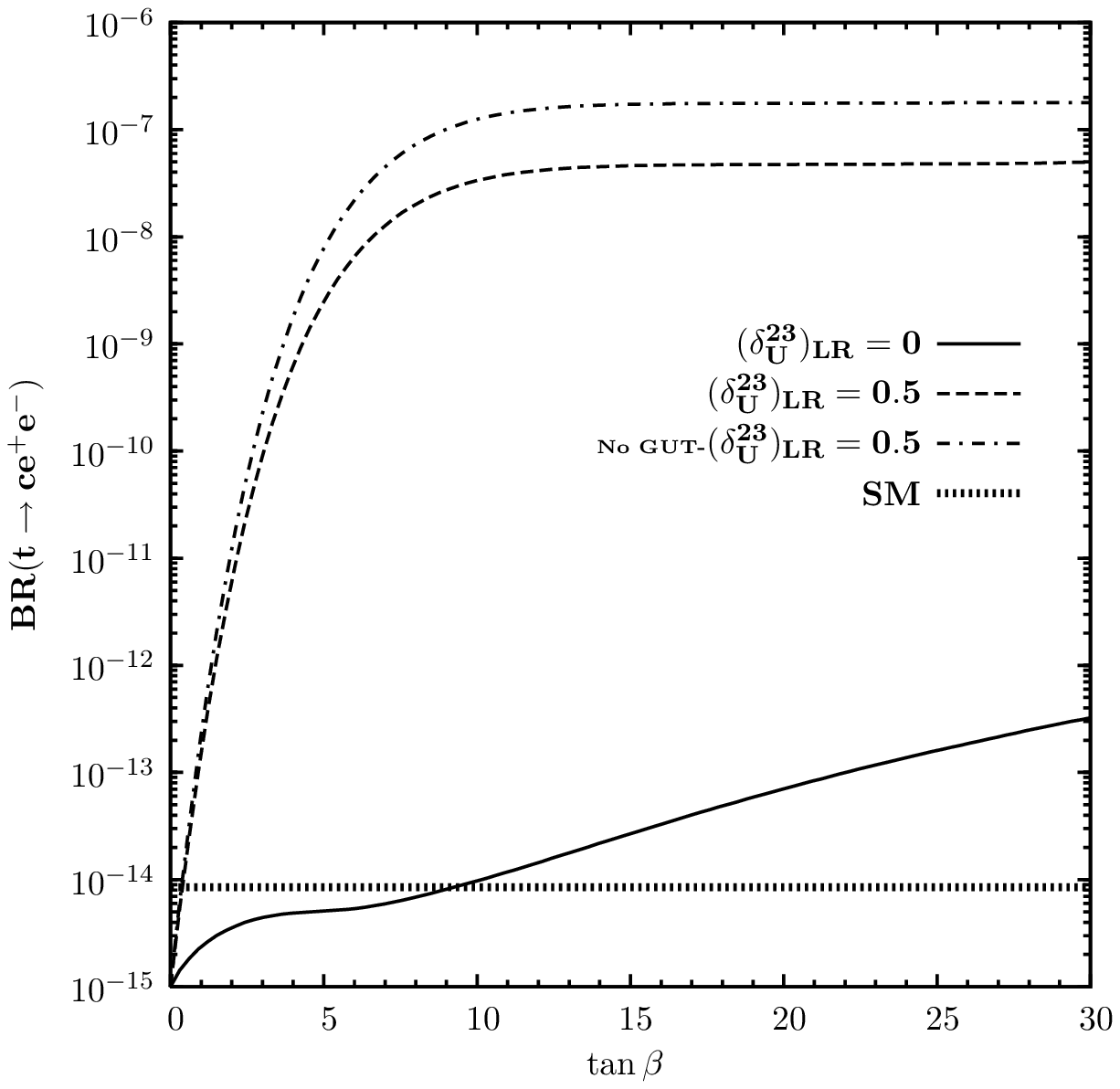}
    \end{array}$
\end{center}
\vskip -0.2in
\caption{
\underline{Left panel}:
BR's of $t\to ce^+e^-$ decay as
function of $M_{\rm {SUSY}}$ for various values
of $(\delta^{23}_U)_{LR}$ with $\tan\beta=10$.
\underline{Right panel}:
BR's as function of $\tan\beta$
under the same conditions with $M_{\rm {SUSY}}=500$ GeV.
The common parameters are chosen as $m_{A^0}=500$ GeV,
$M_2=\mu=200$ GeV, and $A_t=1.2$ TeV. $m_{\tilde{g}}=250$ GeV for
without GUT cases.}\label{fig:tceeMSusyTB}
\end{figure}

We now proceed to investigate the dependence of the branching ratio of $t \to c l^+ l^-$ on the parameters of the supersymmetric model. In the case of flavor violating MSSM, only the mixing between the second and
the third
generations is turned on, and the dimensionless parameters
$\delta$'s run over as much of  the interval
(0,1) as allowed.
The allowed upper limits of $\delta$'s are
constrained by the requirement that
$m_{\tilde{u}_i,\tilde{d}_i}>0$ and  consistent with the experimental
lower bounds (depending on the chosen values of
$M_{\rm {SUSY}},
A, \tan\beta$, and $\mu$).  We
assume a lower bound of $96$ GeV for all up squark masses and $90$ GeV
for the down squark masses \cite{Eidelman:2004wy}. The Higgs
masses are calculated with
\texttt{FeynHiggs version 2.3} \cite{feynhiggs}, with the requirement
that the lightest neutral
Higgs mass is larger than $114$ GeV.  Other experimental bounds
included are  \cite{Eidelman:2004wy}: $96$ GeV for the
lightest chargino,
$46$ GeV the lightest neutralino, and $195$ GeV for the gluino.  We did not include the possible constraints coming from $b \to s \gamma$ or $B_s -{\bar B}_s$ mixing. The reason is the following: the most stringent constraints of these arise from the gluino contributions, and they restrict $(\delta^{23}_D)_{AB}$. Constraints on $(\delta^{23}_U)_{AB}$ are obtained under certain assumptions only. For instance, Cao {\it et. al.} in \cite{Eilam:2006rb}, allow two flavor-violating parameters, $(\delta^{23}_U)_{LL}$ and $(\delta^{23}_U)_{LR}$, to be non-zero at the same time. This permits $(\delta^{23}_U)_{LR}$ to be of order one, but restricts $(\delta^{23}_U)_{LL}$ to be 0.5 or less. 

\begin{figure}[htb]
\begin{center}$
	\begin{array}{cc}
\hspace*{-0.5cm}
	\includegraphics[width=3.1in]{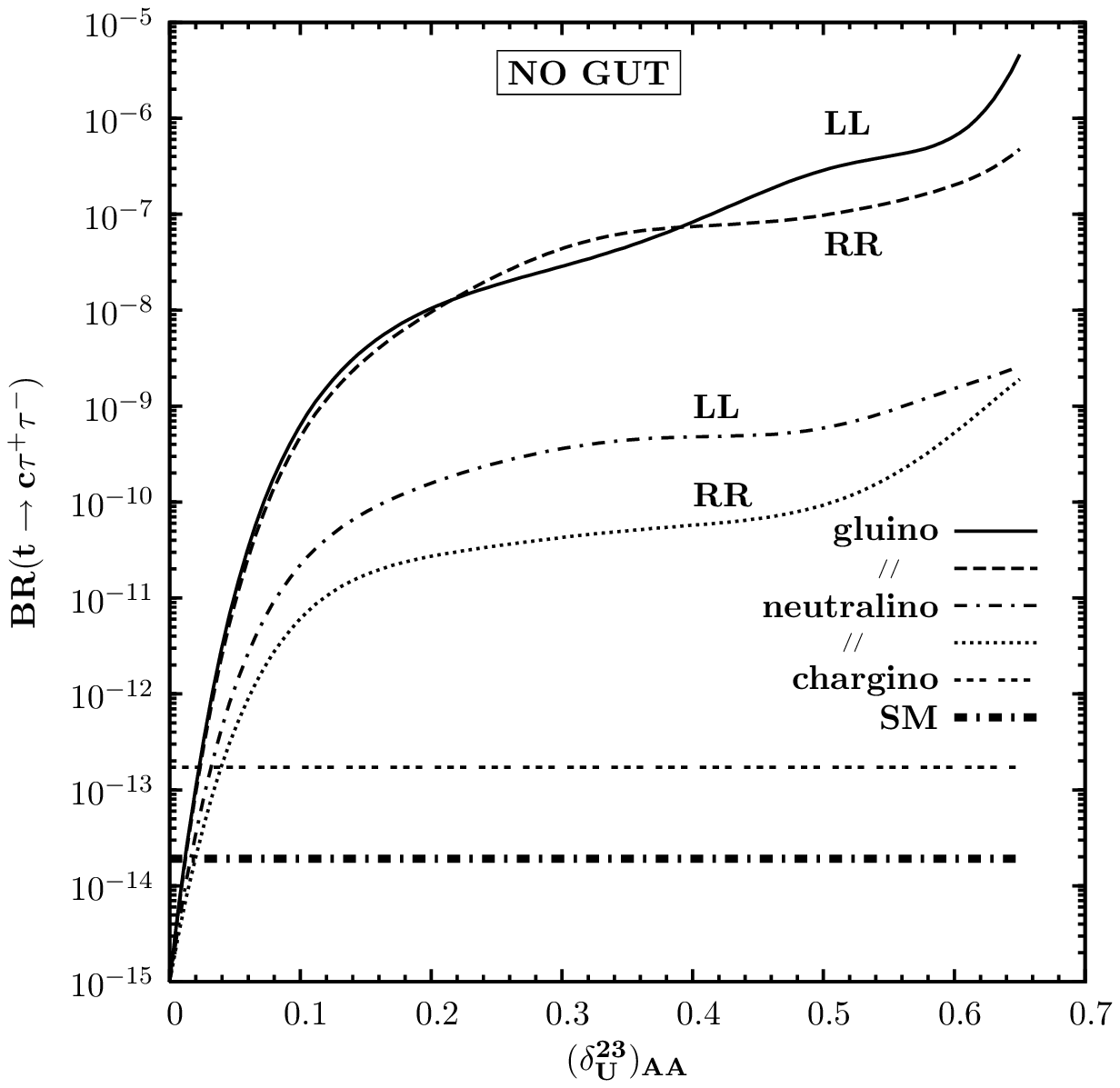} &
	\includegraphics[width=3.1in]{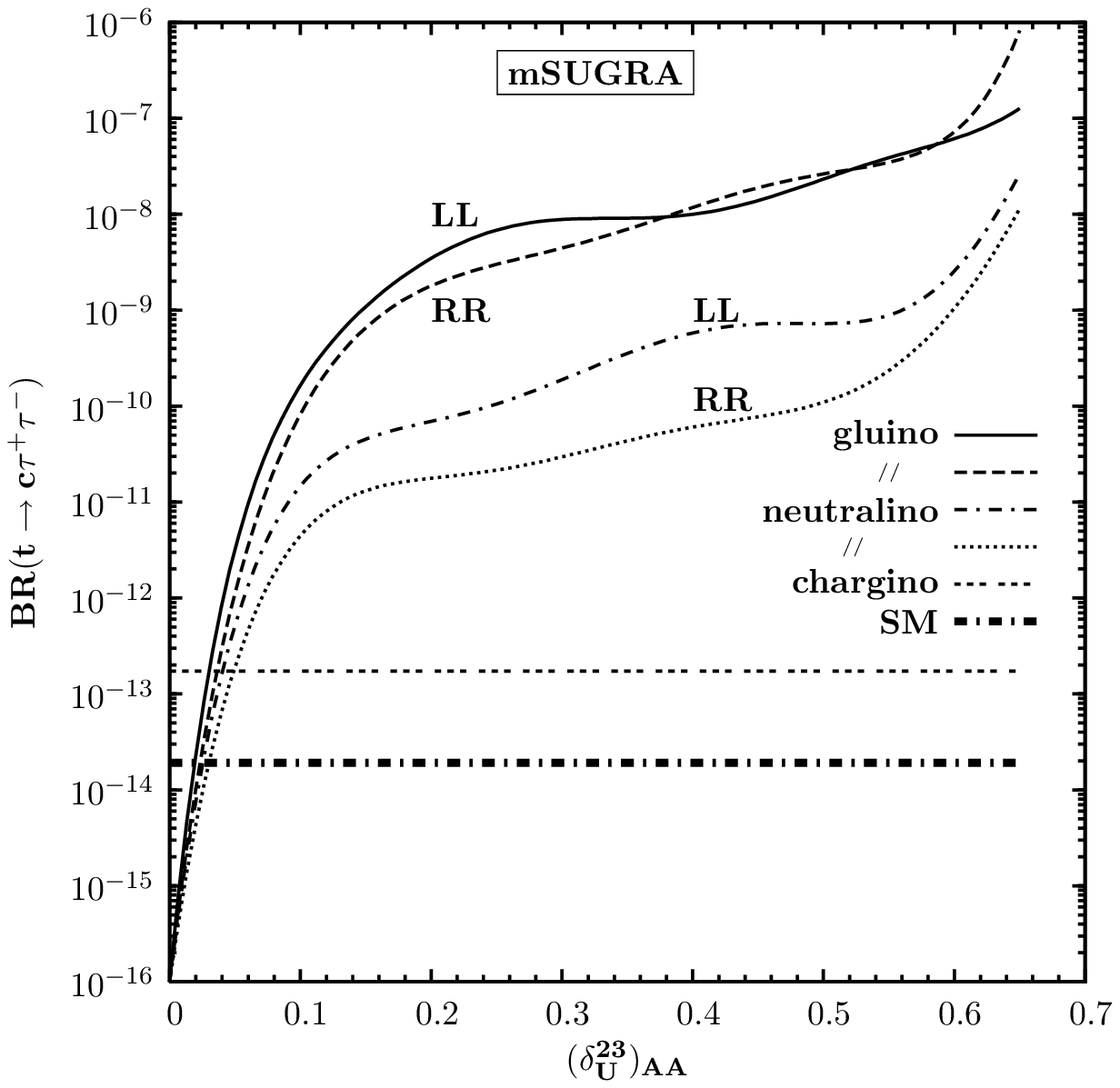} \\
\hspace*{-0.5cm}
	\includegraphics[width=3.1in]{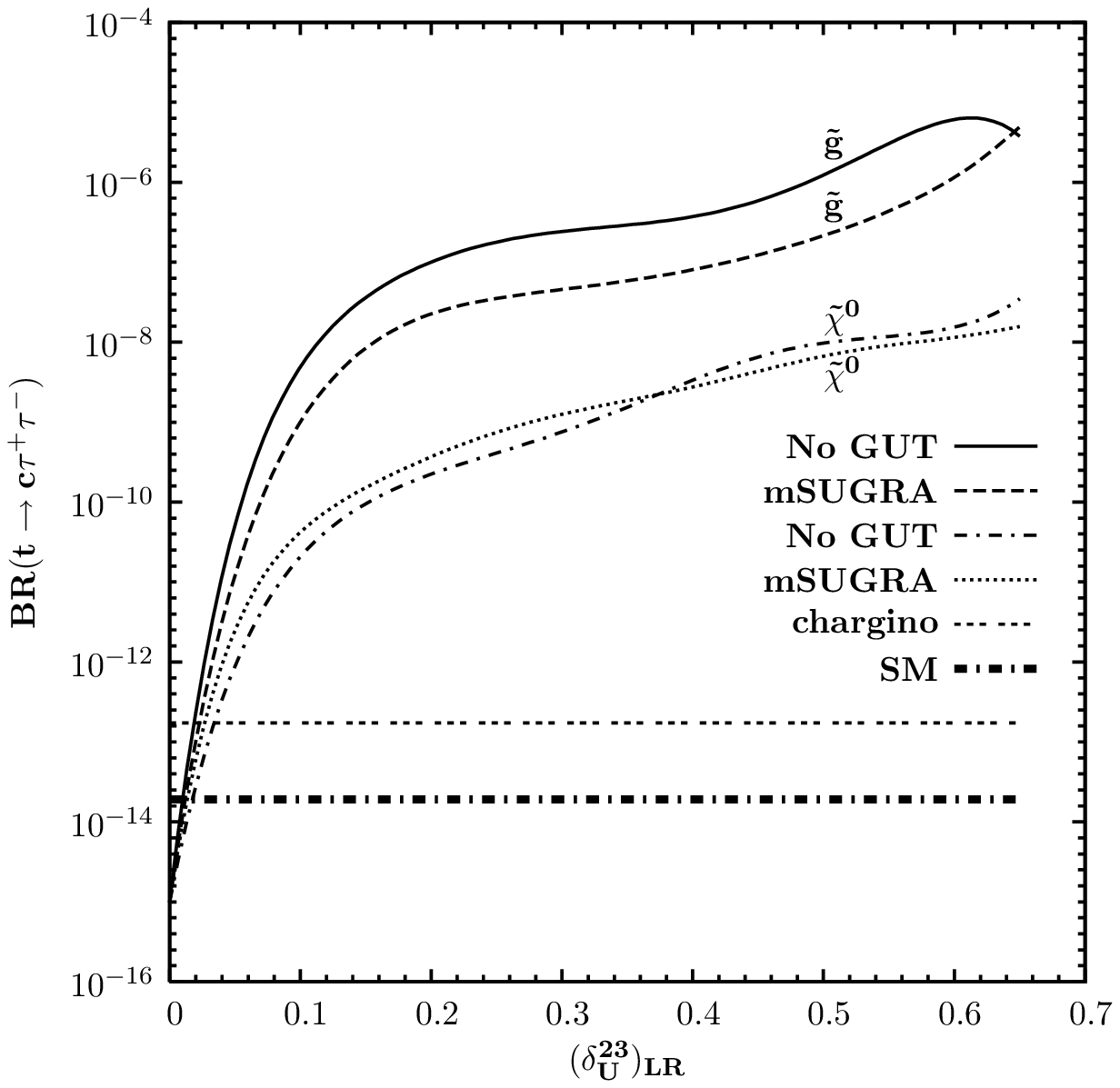} &
	\includegraphics[width=3.1in]{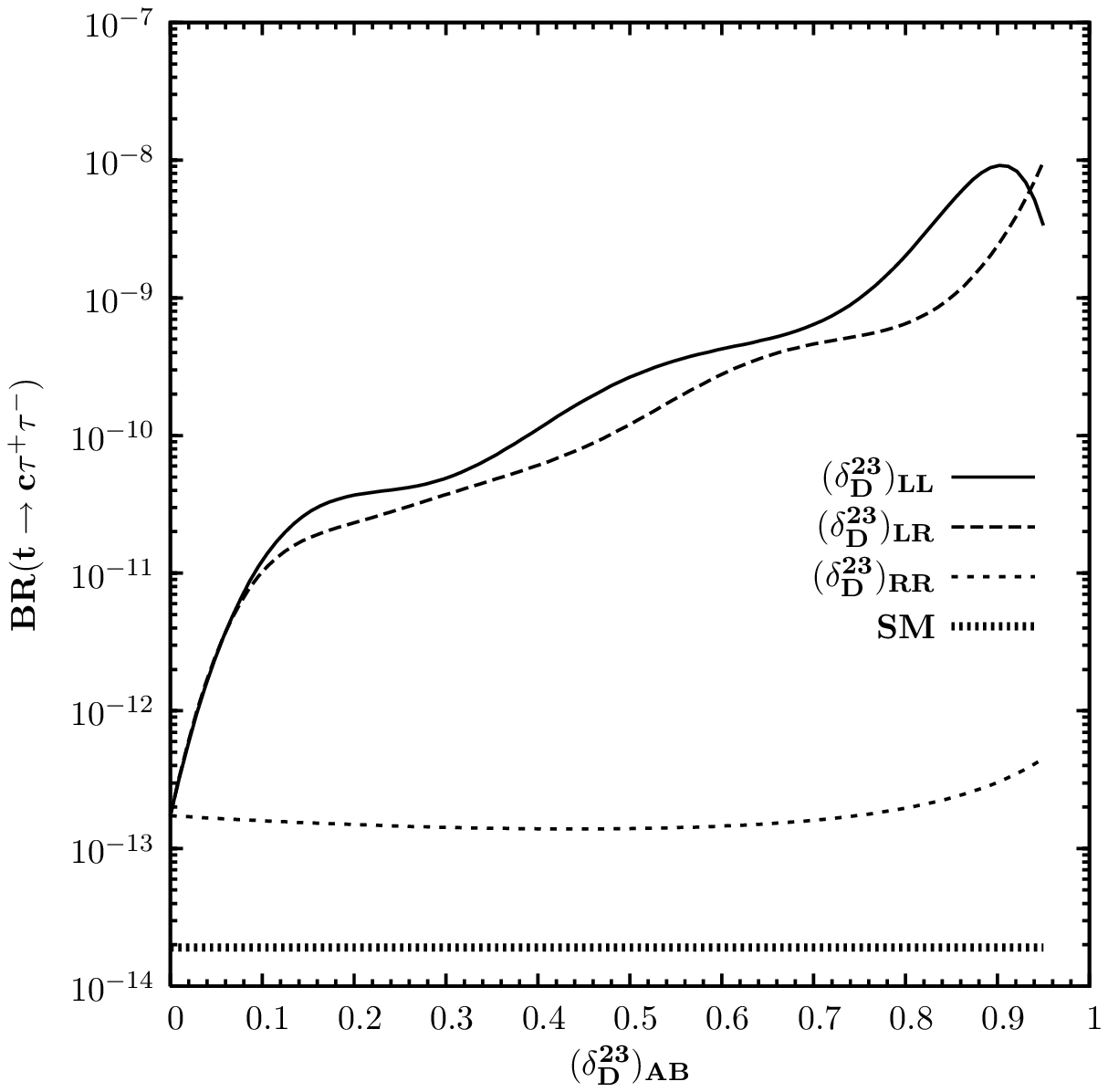}
	\end{array}$
\end{center}
\vskip -0.2in
     \caption{
\underline{Upper left panel}:
BR's of $t\to c\tau^+\tau^-$ decay as
function of $(\delta^{23}_U)_{AA}$ without the GUT relations
($m_{\tilde{g}}=250$ GeV).
\underline{Upper right panel}:
BR's as a function of $(\delta^{23}_U)_{AA}$
under the same conditions in mSUGRA scenario.
\underline{Lower left panel}:
BR's as
function of $(\delta^{23}_U)_{LR}$. 
\underline{Lower right panel}:
BR's as function of $(\delta^{23}_D)_{AB},\,A,B=L,R$.
 The parameters are chosen as $\tan\beta=10$,
$m_{A^0}=M_{\rm {SUSY}}=500$ GeV,
$M_2=\mu=200$ GeV, and $A_t=1.2$ TeV.}\label{fig:tctautauFV}
\end{figure}
\begin{figure}[htb]
\begin{center}$
    \begin{array}{cc}
\hspace*{-0.5cm}
    \includegraphics[width=3.1in]{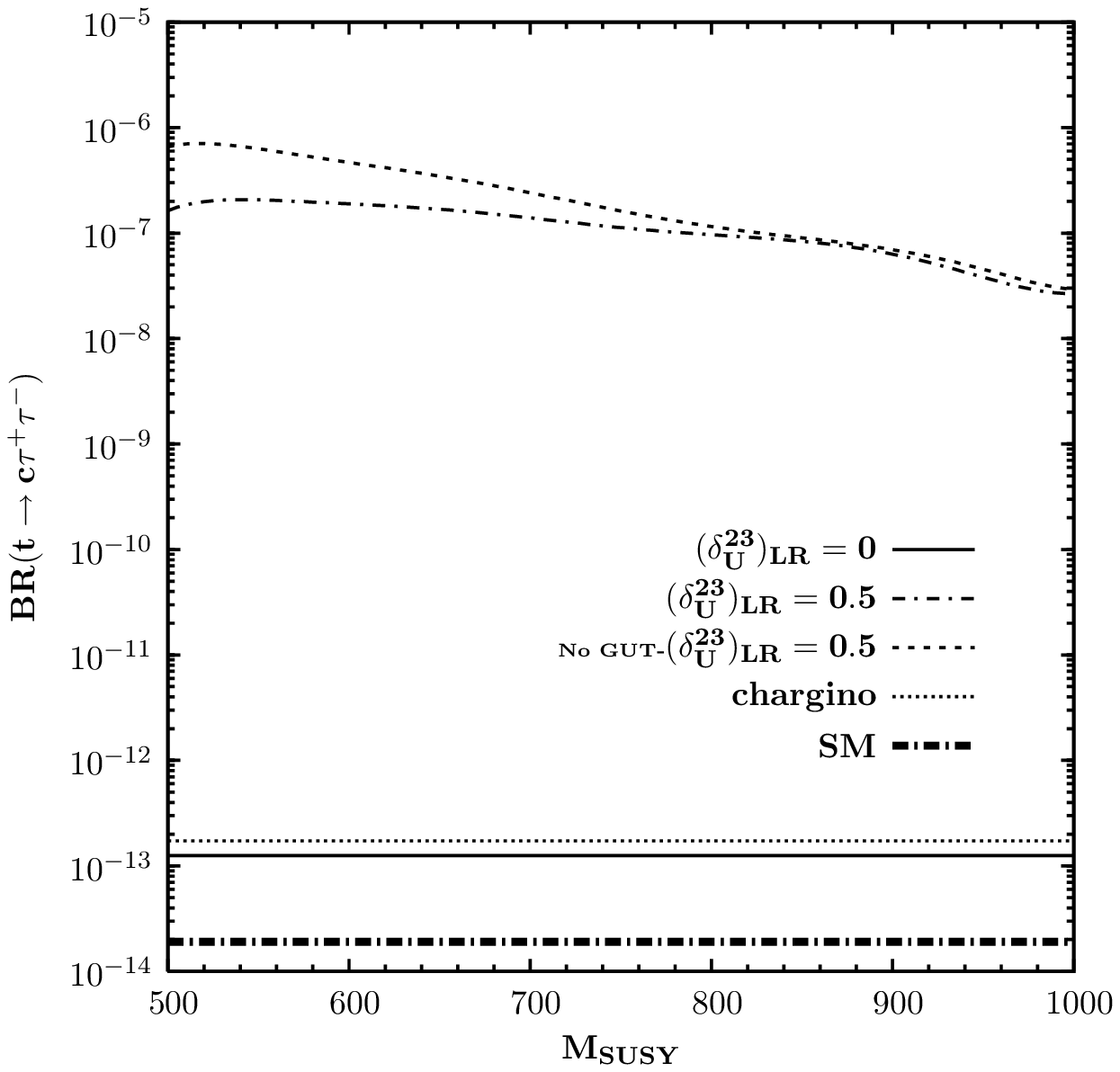} &
    \includegraphics[width=3.1in]{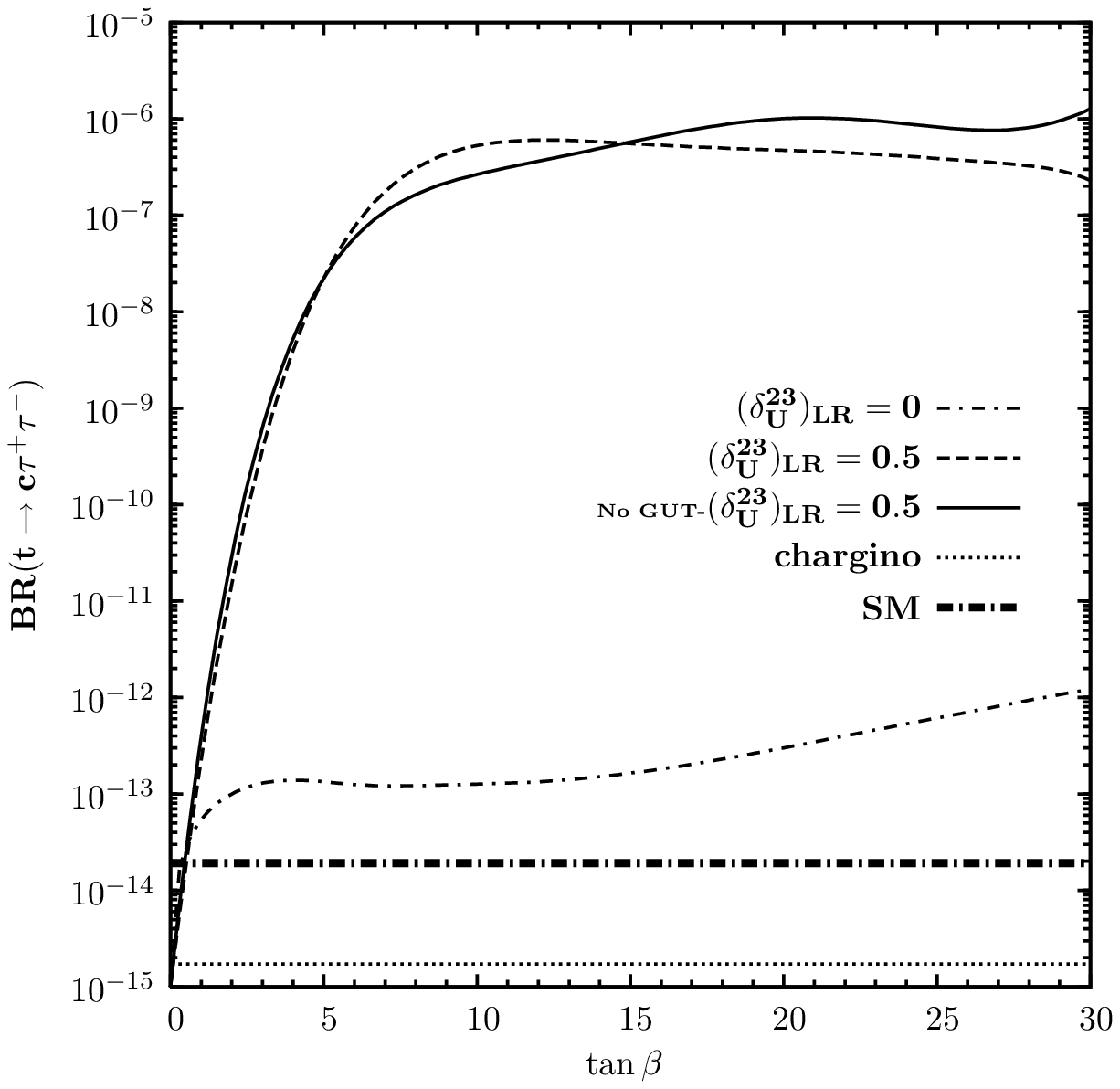}
    \end{array}$
\end{center}
\vskip -0.2in
\caption{
\underline{Left panel}:
BR's of $t\to c \tau^+\tau^-$ decay as
function of $M_{\rm {SUSY}}$ for various values of
$(\delta^{23}_U)_{LR}$ with $\tan\beta=10$.
\underline{Right panel}:
BR's as
function of $M_{\rm {SUSY}}$ for various values of
 $(\delta^{23}_U)_{LR}$ with $\tan\beta=10$.
The common parameters are chosen as $m_{A^0}=500$ GeV,
$M_2=\mu=200$ GeV, and $A_t=1.2$ TeV. $M_{\tilde{g}}=250$GeV for
without GUT cases.}\label{fig:tctautauMSusyTB}
\end{figure}

As in the case of the 2HDM, the constrained MSSM contribution to $t \to c l^+ l^-$ is very small, often below the SM contribution. Thus, the only signals likely to be observed at the colliders would come from the unconstrained MSSM, which we investigate in detail. In Fig.~ \ref{fig:tceeFV} we plot the dependence of the branching ratio of $t \to c e^+ e^-$ on the flavor-changing parameter $\delta_{U,D}^{23}$. The $\delta_U^{23}$ parameter drives the contributions from the gluino and neutralino sectors, while $\delta_D^{23}$ affects the chargino contribution only. The constrained MSSM values for the branching ratio correspond to $\delta_{U,D}^{23}=0$.\footnote{The branching ratio is calculated by taking the top quark total decay width as 1.55 GeV.} The first graph is obtained by allowing  free values for the gaugino masses (no-GUT scenario); in practice this allows the gluino mass to be relatively light and the gluino contribution large. Taking only one $(\delta_U^{23})_{AB}\ne 0,~A,B=L,R$ at a time, the LL and RR contribution from the gluino are almost the same (as the gluino couplings are helicity blind, the only difference comes from the squark masses in the loop). For the neutralino, the LL contribution is larger than the RR one by one order of magnitude or more. The total branching ratio can reach $10^{-7}$ or so, for large allowed values of $(\delta_U^{23})_{LL}$. Once gaugino mass relations are imposed in accordance with the mSUGRA scenario, the gluino masses are forced to be larger and the gluino contributions are reduced by an order of magnitude, while the neutralino contributions are practically unaffected (upper right-sided panel). Throughout both scenarios, the chargino contribution, induced by the CKM matrix, is very small but non-zero. We give, in all graphs, the value of the SM branching ratio, for comparison; but not the value of the branching ration for 2HDM, which is indistinguishable numerically from that of the SM. In the lower left-sided panel we graph the branching ratio of $t \to c e^+ e^-$ as a function of the flavor-  and helicity-changing parameter $(\delta_{U}^{23})_{LR}$. As in the case of $t \to c V$ \cite{Frank:2005vd}, this dependence is somewhat stronger than for the helicity-conserving parameter and the branching ratio can reach $10^{-6}$ for the no-GUT scenario. Finally in the lower right-sided panel, we show the dependence of he branching ratio of $t \to c e^+ e^-$ as a function of the  parameter $(\delta_{D}^{23})_{AB}$ in the down squark sector. This parameter affects the chargino contributions only. The branching ratio can reach at most $10^{-9}$ from the chargino contributions alone; in this case $(\delta_{U}^{23})_{AB}$ is set to zero and the gluino and neutralino contributions are zero. Should the flavor violation come from the down squark sector only, the branching ratio would be one-to-two orders of magnitude smaller than if it originated in the up squark sector. 

In Fig.~\ref{fig:tceeMSusyTB} we plot the dependence of the branching ratio of $t \to c e^+e^-$ on the scalar mass $M_{\rm SUSY}$ and $\tan \beta$. $M_{\rm SUSY}$ affects scalar quark masses; one can see from the left-handed panel in Fig.~\ref{fig:tceeMSusyTB} that the dependence is very weak, less than one order of magnitude change in going from 500 GeV to 1 TeV. The same is true for the $\tan\beta$ dependence which affects scalar quark masses and mixings, and the chargino/neutralino contribution: except for very low $\tan \beta$'s, between 0 and 5 when the branching ratio increases dramatically from 0, the branching ratio is insensitive to variations in intermediate values of $\tan \beta$ between 7 and 30. 

We investigated the same dependence for the decay $t \to c \mu^+ \mu^-$ and the results are practically identical, so we do not show them here. However, as expected, the branching ratio for $t \to c \tau^+ \tau^-$ differs numerically, though not in its general variation pattern,  from $t \to c e^+ e^-$ and we show it in Fig.~\ref{fig:tctautauFV}. While the differences between the behavior of different curves are insignificant, the branching ratio for $t \to c \tau^+ \tau^-$ is consistently  one order of magnitude larger than either $t \to c e^+ e^-$ or $t \to c \mu^+ \mu^-$. Plotting the branching ratio of $t \to c \tau^+ \tau^-$
with respect to $M_{\rm SUSY}$ and $\tan \beta$, in Fig. \ref{fig:tctautauMSusyTB}, one can see the effects  of the $\tau$ lepton coupling to Higgs. In this, we assume flavor-violation in the up squark sector only, thus only the gluino and neutralino graphs are important.  Looking at the diagrams, the decays proceed mostly though the penguins  $t \to c \gamma^*$, $t \to c Z^*$ and $t \to c H^*$ followed by $\gamma^*, Z^*, H^* \to \tau^+ \tau^-$. If the Higgs exchange was dominant, one would expect an enhancement in $\tau \tau$ with respect to $ee$ final state, of the order of $(m_\tau/m_e)^2$, while no change would be expected if the dominant exchange would be through the $\gamma$ or $Z$.\footnote{Note that in the $\tau\tau$ case there is an additional phase space supression.} The enhancement shows that there is some interference between the Higgs and vector boson graphs.

\section{$t \to c \nu_i {\bar \nu}_i$}
\begin{figure}[htb]
        \centerline{\epsfxsize 6.5in {\epsfbox{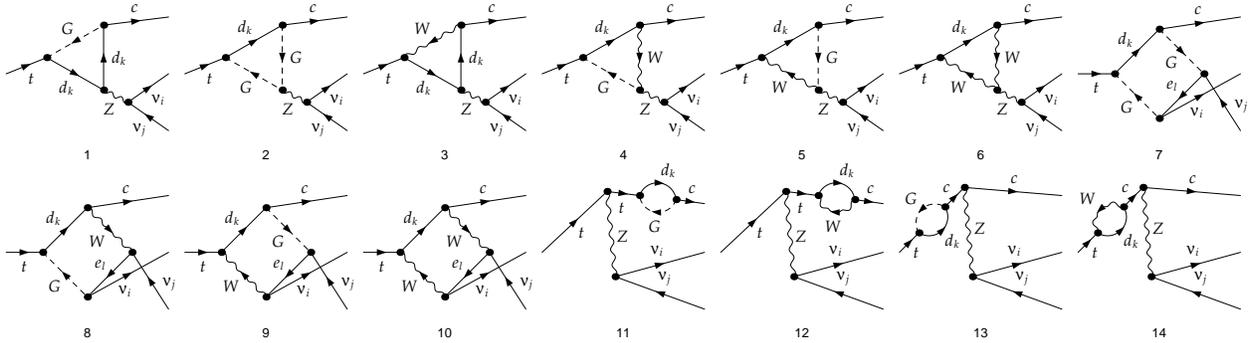}}}
\caption
      {The one-loop SM contributions to $t\to c \nu_i {\bar \nu}_i$ in the
't Hooft-Feynman gauge.}
\label{fig:SM_nunu}
\end{figure}

\begin{figure}[htb]
        \centerline{ \epsfxsize 6.5in {\epsfbox{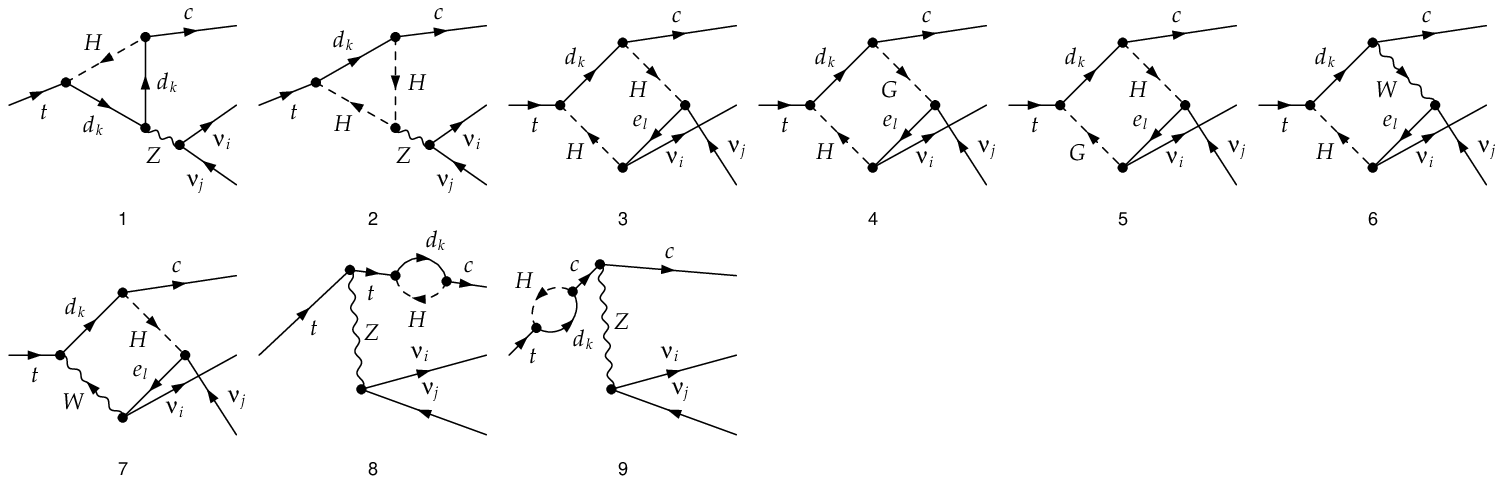}}}
\caption
      {The one-loop 2HDM contributions to $t\to c \nu_i{\bar \nu}_i$ in the
't Hooft-Feynman gauge.}
\label{fig:THDM_nunu}
\end{figure}

\begin{figure}[htb]
        \centerline{\epsfxsize 4in {\epsfbox{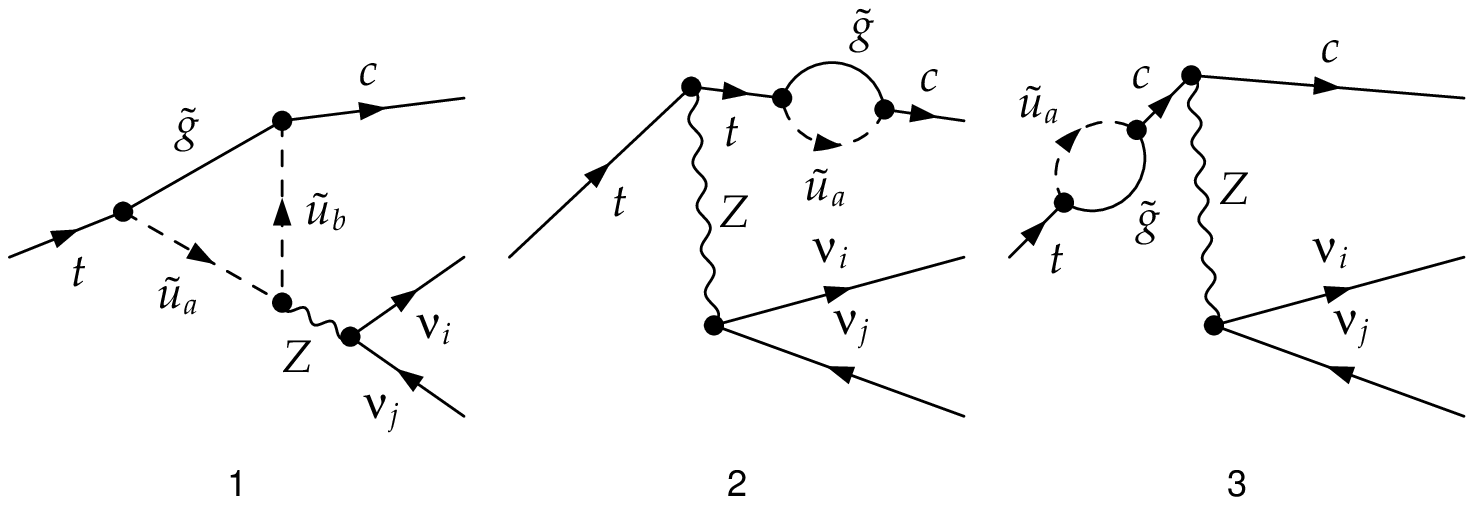}}}
\caption
      {The one-loop gluino contributions to $t\to c \nu_i\bar{\nu}_i$ in MSSM in the
't Hooft-Feynman gauge.}
\label{fig:gluino_nunu}
\end{figure}

\begin{figure}[htb]
        \centerline{\epsfxsize 6.5in {\epsfbox{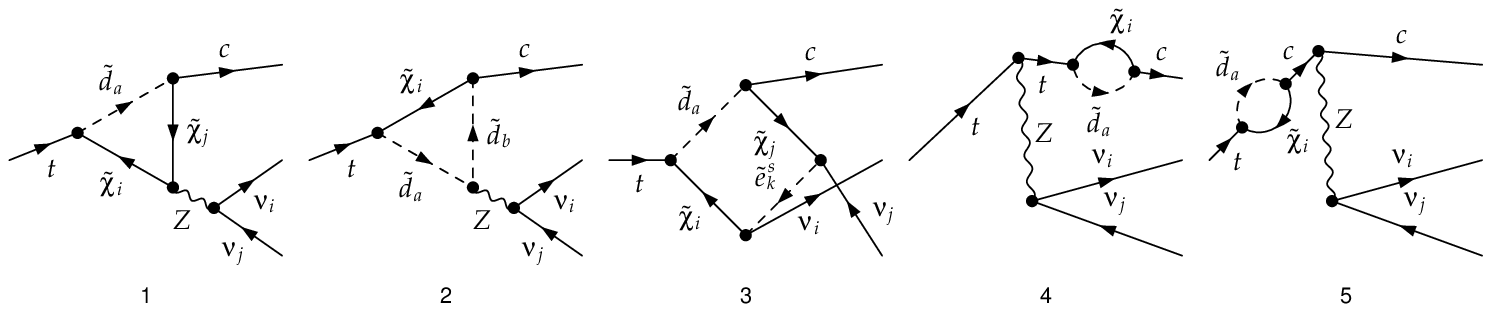}}}
\caption
      {The one-loop chargino contributions to $t\to c \nu_i{\bar \nu}_i$ in MSSM in the
't Hooft-Feynman gauge.}
\label{fig:chargino_nunu}
\end{figure}

\begin{figure}[htb]
        \centerline{\epsfxsize 6.5in {\epsfbox{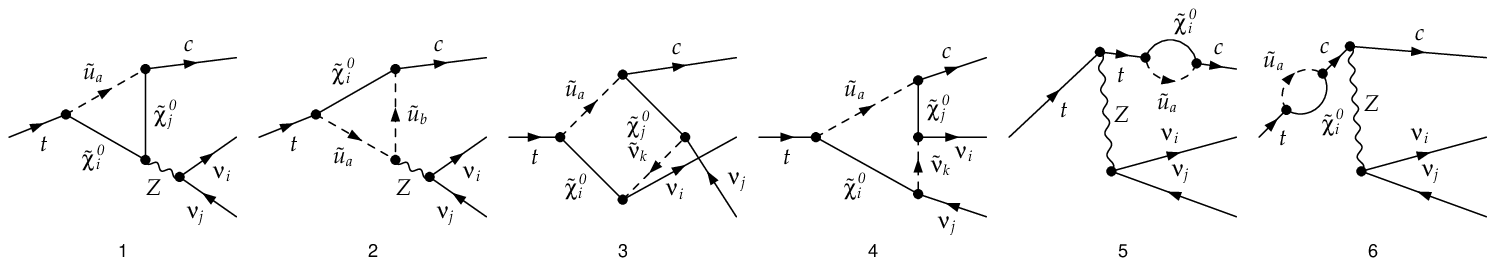}}}
\caption
      {The one-loop neutralino contributions to $t\to c \nu_i\bar{\nu}_i$ in MSSM in the
't Hooft-Feynman gauge.}
\label{fig:neutralino_nunu}
\end{figure}

The transition $t \to c \nu_i {\bar \nu}_i$ is generated at one-loop by the same operators  as $t \to c l^+ l^-$. Of course, in this case, the experimental search for this rare decay is harder than for $t \to c l^+ l^-$ (the signal will be a quark jet plus missing energy instead of a quark jet plus a lepton pair). As one cannot distinguish neutrino species, we sum over all three light families, which in principle could result in a rate for $t \to c \sum \nu_i \bar{\nu}_i$ a factor of 3-4 larger than the rate for $t \to c l^+ l^-$. The Feynman diagrams which contribute to this decay in  the SM are given in Fig. \ref{fig:SM_nunu}; for the 2HDM in Fig \ref{fig:THDM_nunu}; and for the gluino, chargino and neutralino contributions in MSSM in Figs. \ref{fig:gluino_nunu}, \ref{fig:chargino_nunu} and \ref{fig:neutralino_nunu} respectively.
\begin{figure}[htb]
\begin{center}$
	\begin{array}{cc}
\hspace*{-0.5cm}
	\includegraphics[width=3.1in]{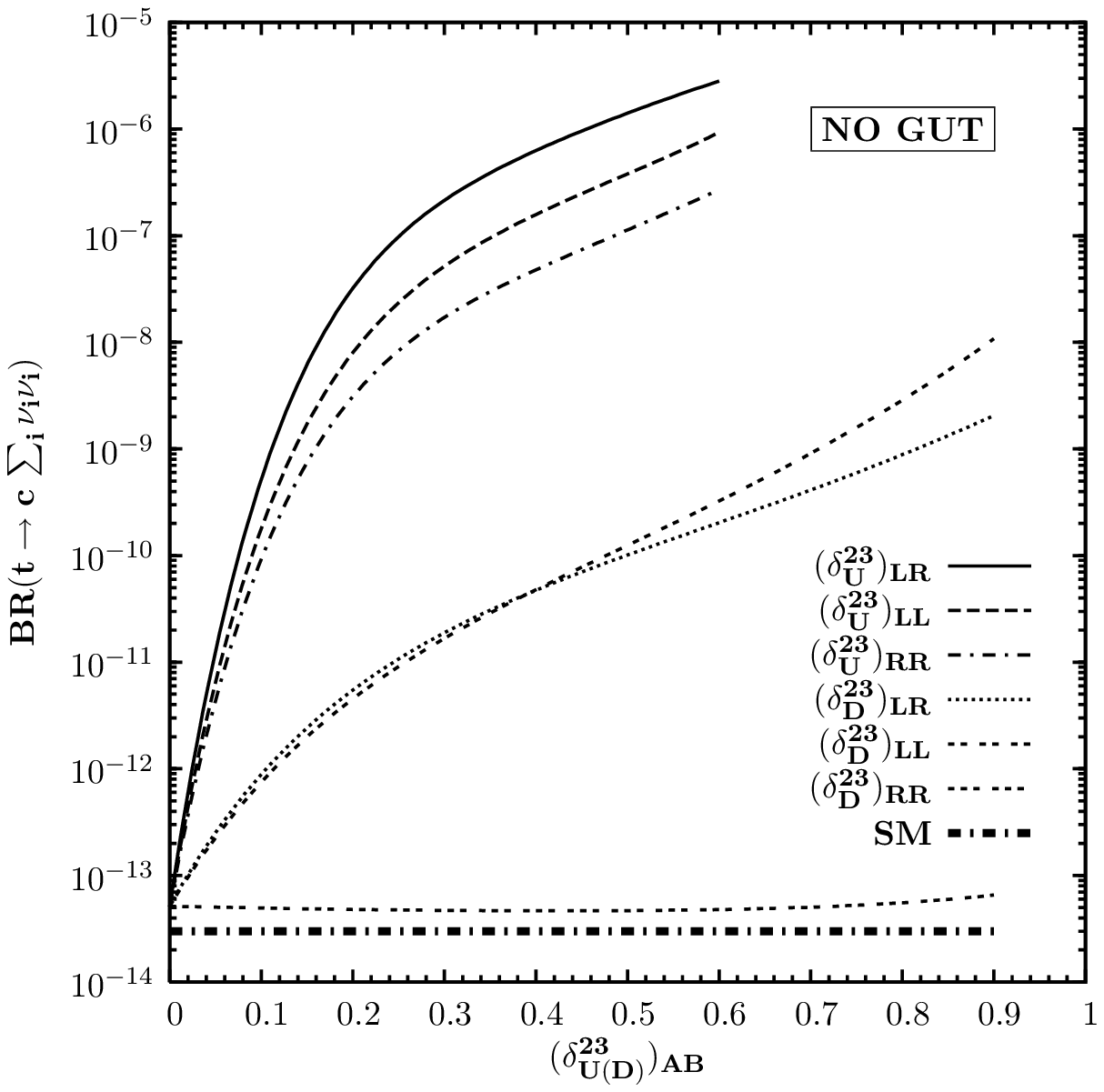} &
	\includegraphics[width=3.1in]{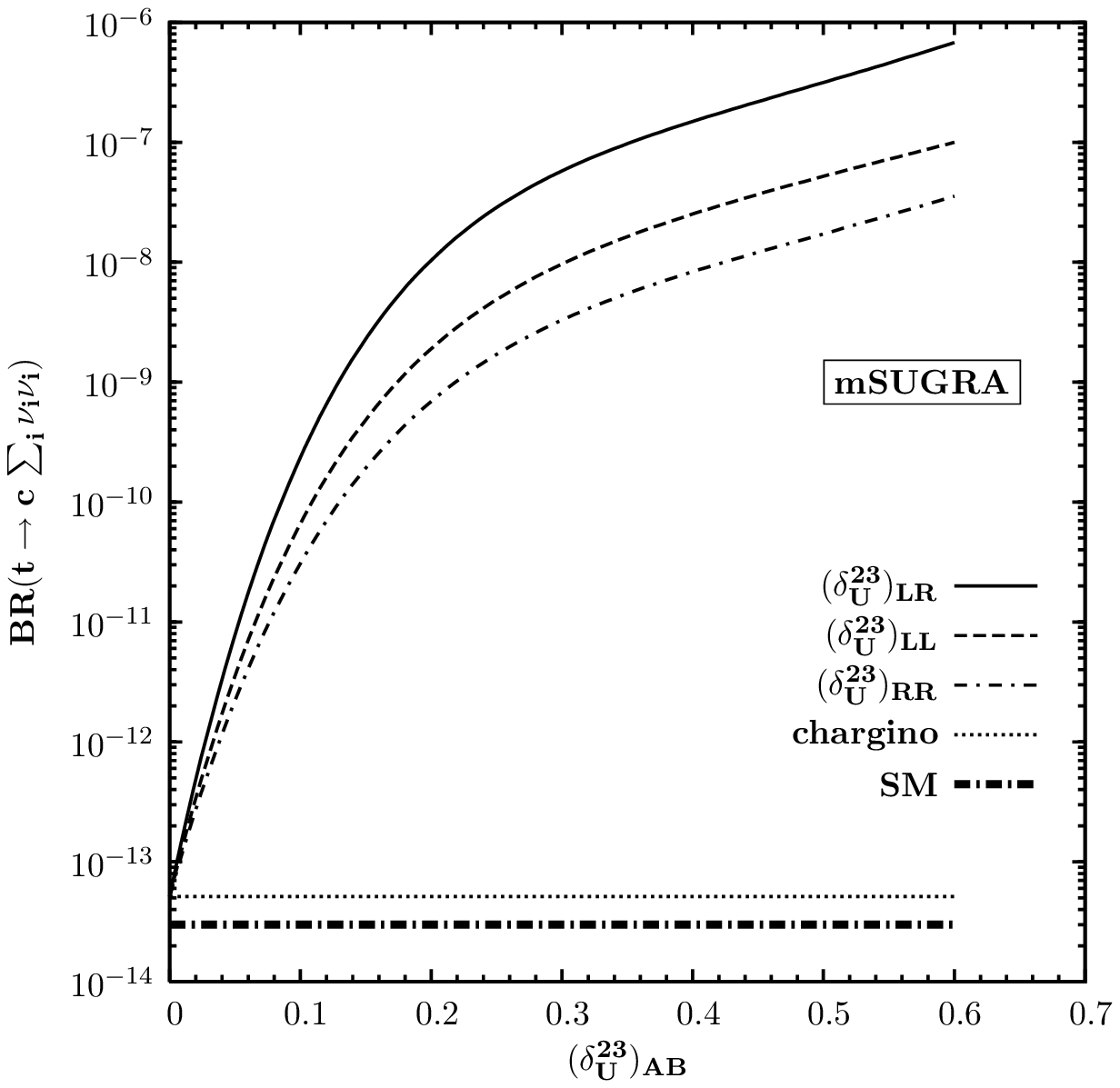} \\
\hspace*{-0.5cm}
	\includegraphics[width=3.1in]{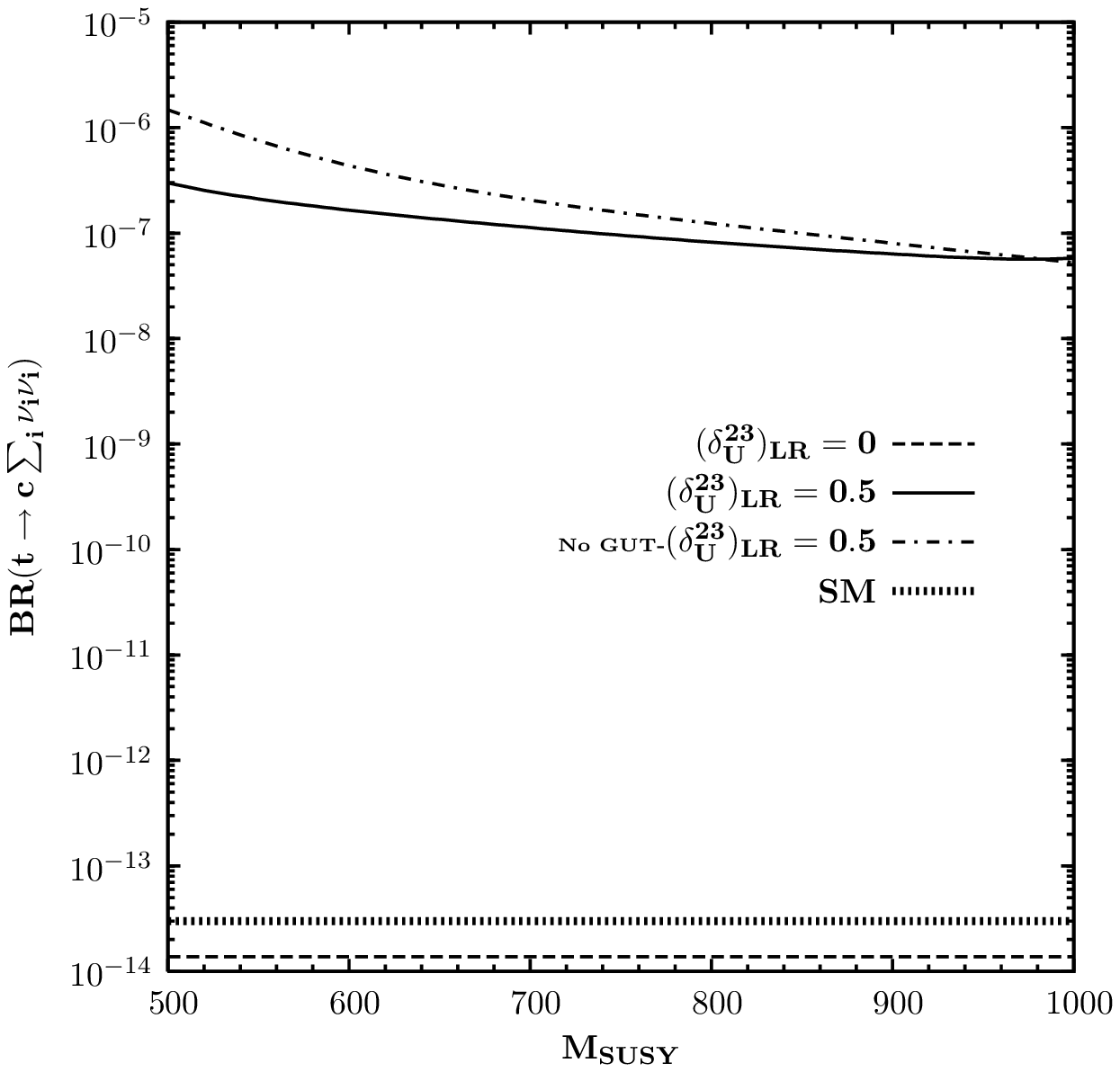} &
	\includegraphics[width=3.1in]{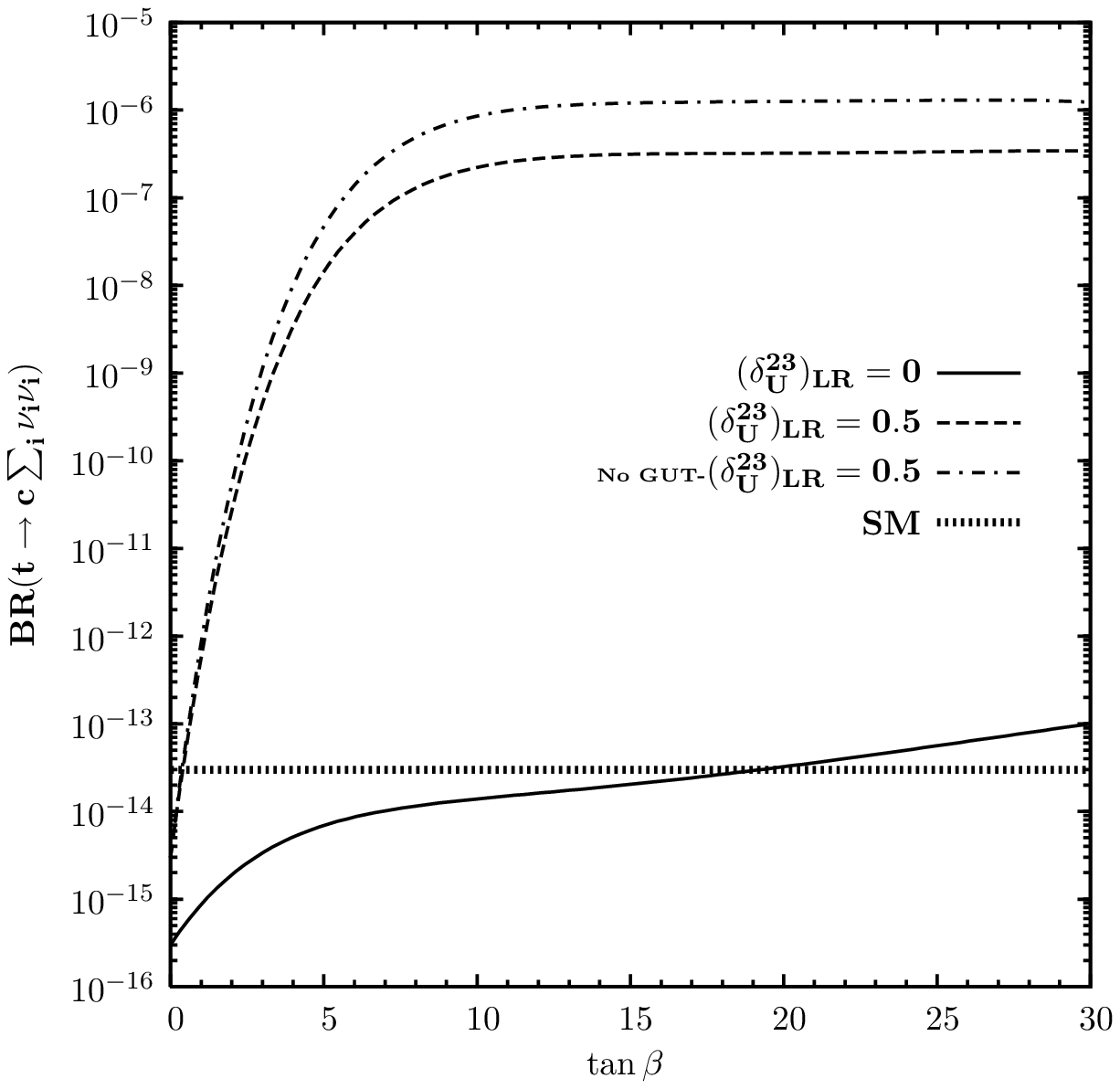}
	\end{array}$
\end{center}
\vskip -0.2in
\caption{
\underline{Upper left panel}:
BR's of $t\to c\sum_i \nu_i{\bar \nu}_i$ decay as
function of $(\delta^{23}_{U(D)})_{AB}$ without the
GUT relations ($m_{\tilde{g}} =250$ GeV).
\underline{Upper right panel}:
BR's as function of $(\delta^{23}_U)_{AB}$
under the same conditions in mSUGRA scenario.
The parameters are chosen as $\tan\beta=10$,
$m_{A^0}=M_{\rm {SUSY}}=500$
GeV, $M_2=\mu=200$ GeV, and $A_t=1.2$ TeV.
\underline{Lower left panel}:
BR's as
function of $M_{\rm {SUSY}}$ for various values of 
$(\delta^{23}_U)_{LR}$ for various values of
$(\delta^{23}_U)_{LR}$ with $\tan\beta=10$.
\underline{Lower right panel}:
BR's as a function of $\tan\beta$ under the
same conditions with $M_{\rm {SUSY}}=500$ GeV.}\label{fig:tcnunuAB}
\end{figure}

Fig. \ref{fig:tcnunuAB} 
is dedicated to the investigation of the branching ratio of $t \to c \sum \nu_i {\bar \nu}_i$. For the case in which the gluino mass is allowed to be relatively light (no GUT scenario), the largest contribution comes from the flavor- and helicity-changing parameter in the up squark sector $(\delta_{U}^{23})_{LR}$, though the contributions from the LL and RR parameters are comparable. We recover here the features from the decay $t \to c l^+ l^-$, while summing over neutrino flavors results in a larger overall branching ratio than $t \to ce^+ e^-$ (The factor is slightly bigger than three). The down sector parameters $(\delta_{D}^{23})_{AD}$ contribute much less, especially $(\delta_{D}^{23})_{RR}$ as there are no right-handed gauginos. At best, the branching ratio for $t \to c \sum \nu_i \bar{\nu}_i$ can reach $10^{-6}$. This decreases somewhat for the case where we impose GUT relations between gaugino masses (upper right-handed panel). We plot the variation with $(\delta_{U}^{23})_{AB}$ parameters only, for comparison, as they are dominant and the pattern of variation remains the same. The variation with the scalar mass $M_{\rm SUSY}$ and $\tan \beta$ are shown in the lower right-had panel. As for $t \to c l^+ l^-$, the variation with $M_{\rm SUSY}$ is weak, the branching ratio decreasing by less than one order of magnitude when varying  $M_{\rm SUSY}$ from 500 GeV to 1 TeV. Also, the variation with $\tan \beta$ is very pronounced for low $\tan \beta$ (between 0 and 5) but the branching ratio is insensitive to variations in intermediate values of $\tan \beta$ between 10 and 30. 

\section{Production of single top in $e^+e^- \to t {\bar c}$ at the ILC}

While the LHC as a factory of top quarks would allow to search for FCNC top quark decays, the single top quark production is also likely to be overwhelmed by the large background. At the ILC at $\sqrt{s} \le 500$ GeV, the signal $e^+e^- \to t {\bar c}$ is likely to be observed as a clear signal at relatively low energy, $ \le 2m_t$.  At such energies a single top quark production would present a clear signal, with the $\bar c$ being essentially a massless jet \cite{Agashe:2006wa}. This production process has been considered by Chang {\it et al} \cite{Chang:1992qa} in the SM and its extensions, then 
by Huang {\it et al} \cite{Huang:1999bt} in the SM, and later by Li {\it et al} \cite{Li:1999ms} in MSSM including SUSY-QCD corrections only.  Lately, the process has been reconsidered by Cao {\it et al} \cite{Cao:2002si} in the unconstrained MSSM (SUSY QCD) with a nonzero mixing $(\delta_{U}^{23})_{LL}$. The conclusion is that the production in SM is impossible to observe but it could be observable in models beyond the SM. Our results agree with \cite{Cao:2002si} for the same parameter values. Here,  the charm associated top quark production at ILC, $e^+e^- \to t {\bar c}$, is discussed in both constrained and unconstrained MSSM frameworks including gluino, chargino, neutralino and charged Higgs contributions, allowing flavor violation in both up and down squark sectors between second and third generation. Additionally, we calculate the forward-backward asymmetry. We assume that the electrons and positrons are unpolarized.
\begin{figure}[htb]
\begin{center}$
	\begin{array}{cc}
\hspace*{-0.5cm}
	\includegraphics[width=3.1in]{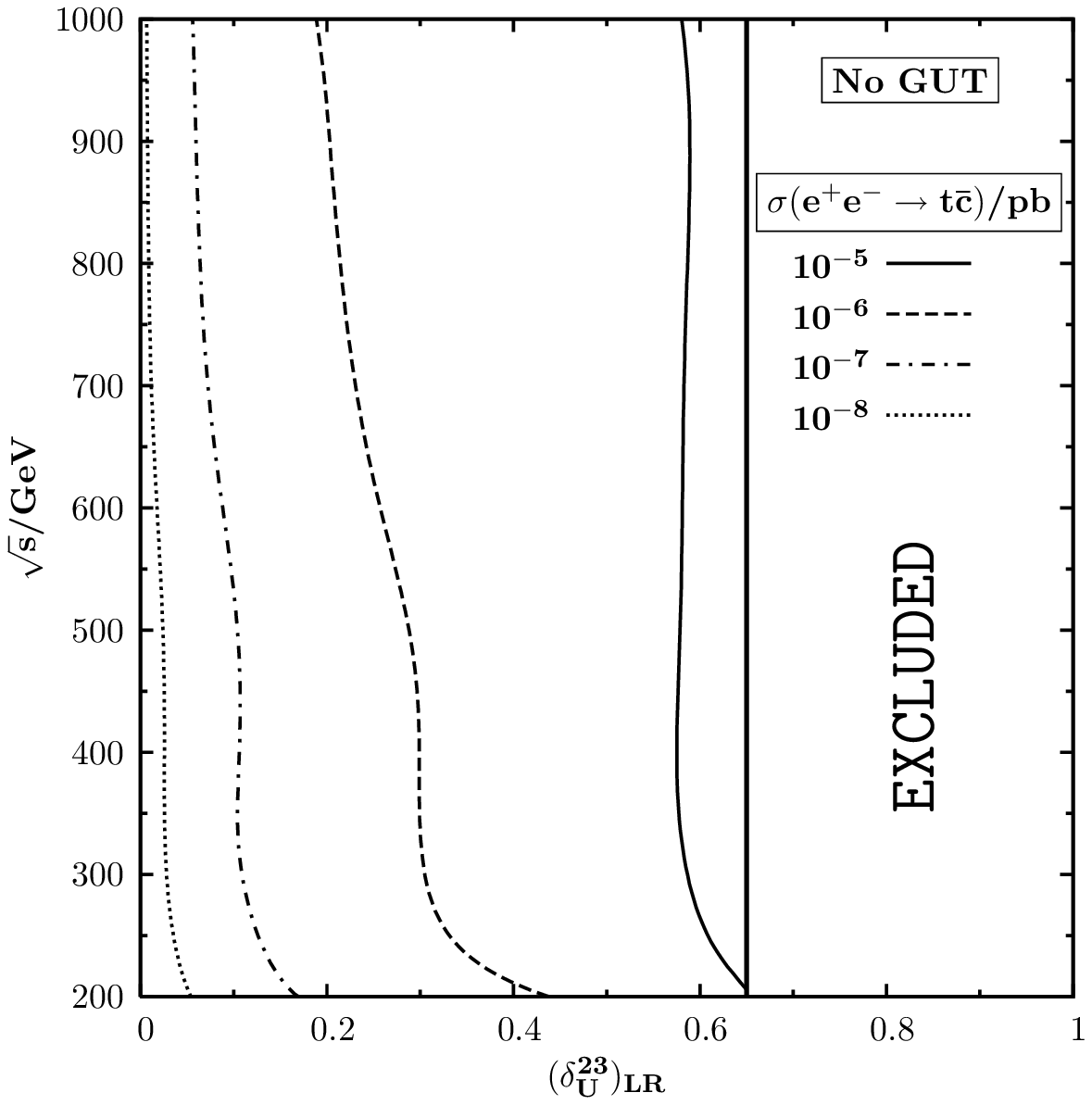} &
	\includegraphics[width=3.1in]{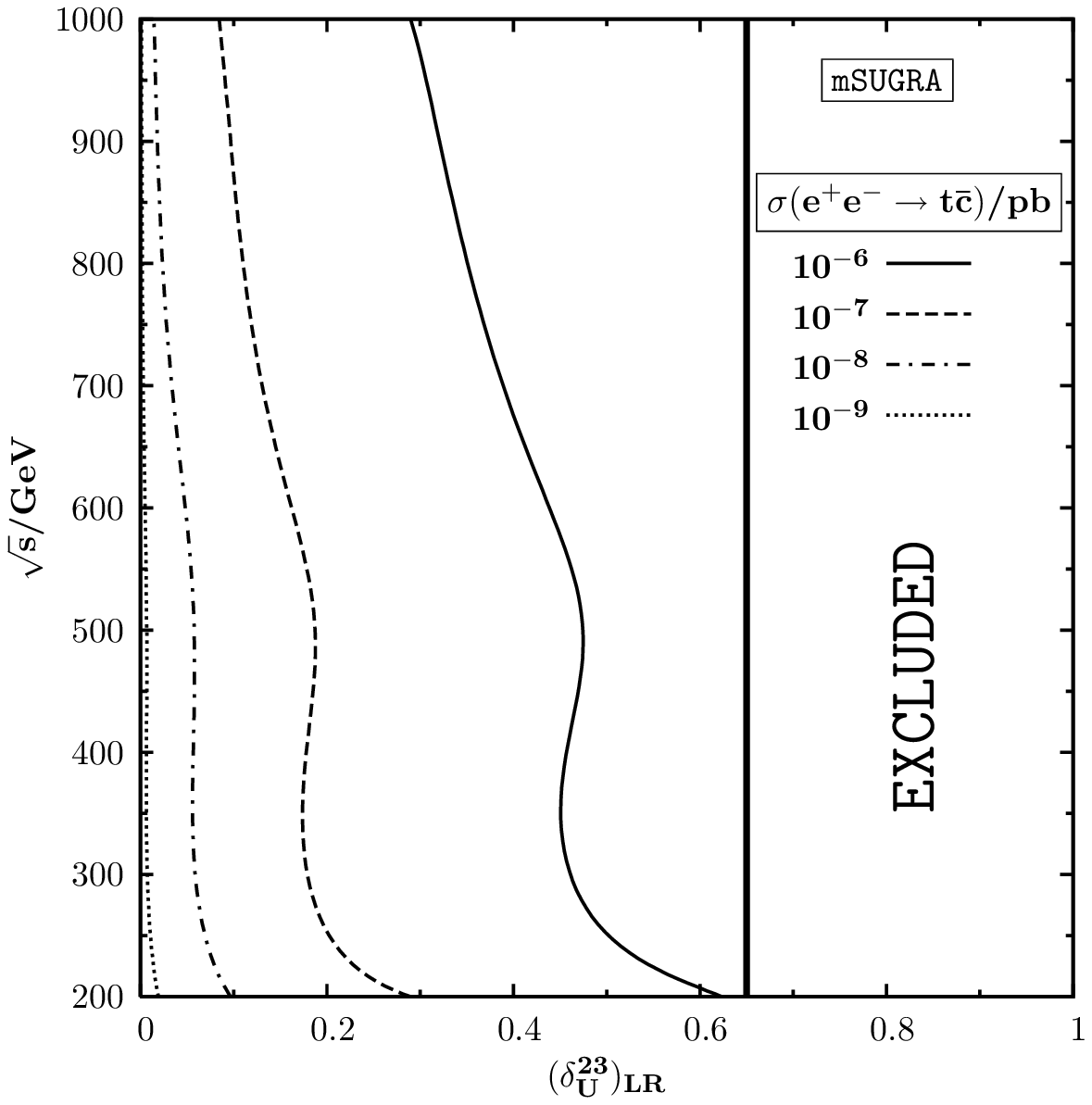} \\
\hspace*{-0.5cm}
	\includegraphics[width=3.1in]{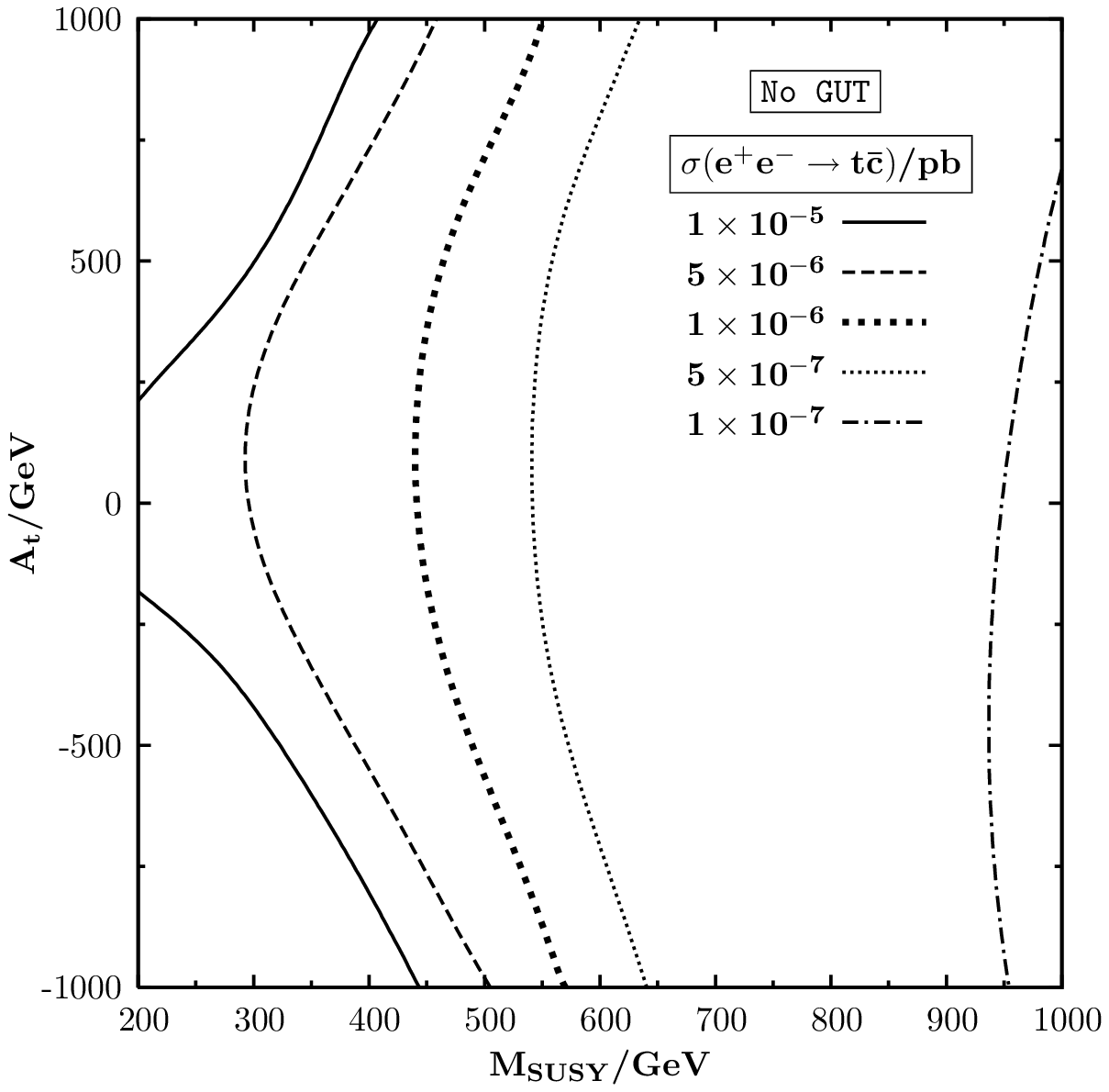} &
	\includegraphics[width=3.1in]{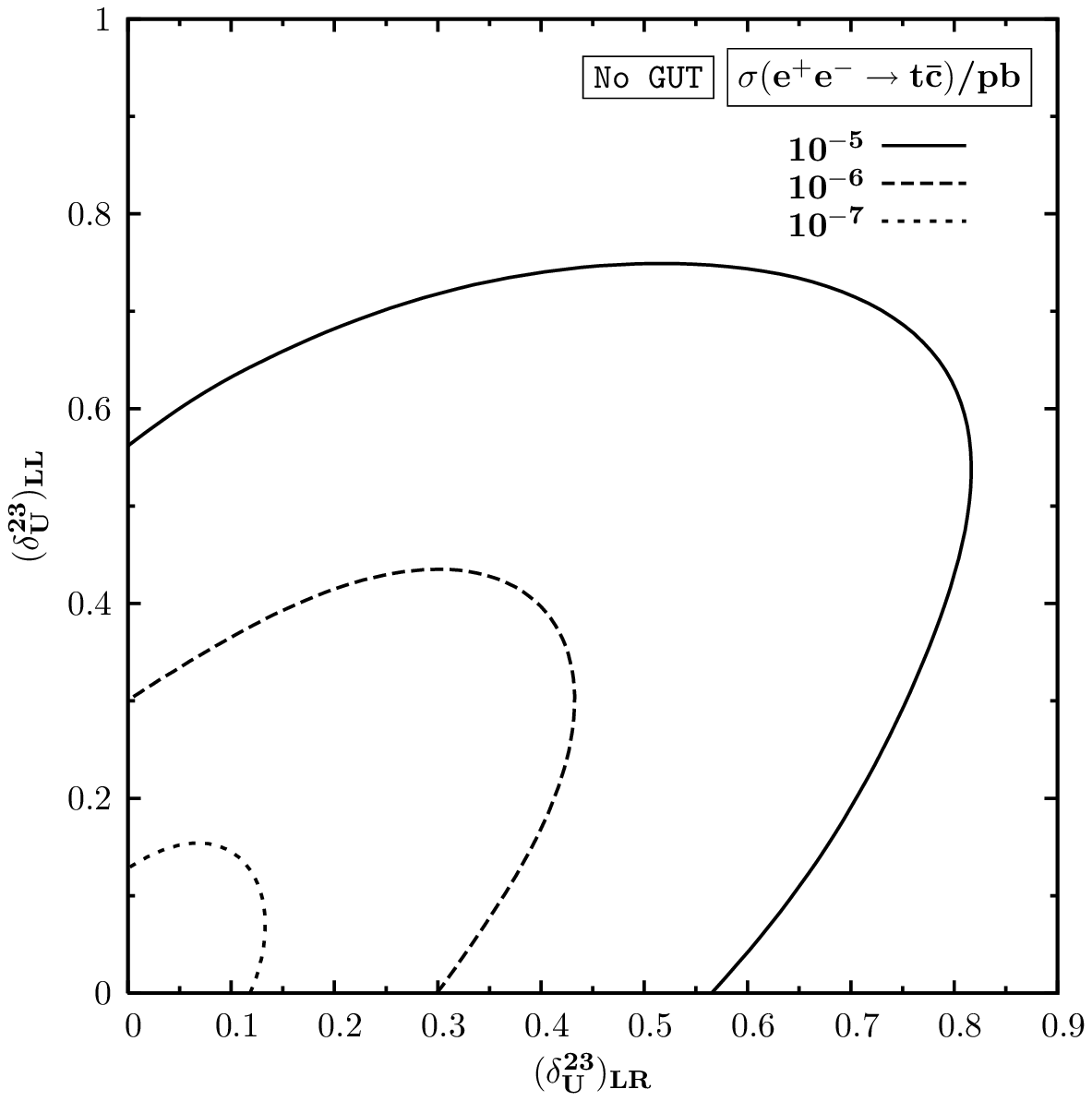} \\
	\end{array}$
\end{center}
\vskip -0.2in
\caption{
\underline{Upper left panel}:
The cross section of $e^+e^-\to t\bar{c}+{\bar t}c$ in 
2-dimensional contour plot
without the
GUT relations ($m_{\tilde{g}} =250$ GeV).
\underline{Upper right panel}:
The cross section in 
2-dimensional contour plot
in mSUGRA scenario. The excluded regions are
determined by the positivity of squark masses. 
The parameters are chosen as $\tan\beta=10$,
$m_{A^0}=M_{\rm {SUSY}}=500$
GeV, $M_2=\mu=200$ GeV, and $A_t=1.2$ TeV.
\underline{Lower left panel}:
The cross section in 
2-dimensional contour plot
without GUT relations ($m_{\tilde{g}} =250$ GeV) 
with 
$(\delta_U^{23})_{LR}=0.5$.
\underline{Lower right panel}:
The cross section in 
2-dimensional contour plot
without GUT relations ($m_{\tilde{g}} =250$ GeV) 
under the same parameter values.
}\label{fig:CS}
\end{figure}

\begin{figure}[htb]
\begin{center}$
	\begin{array}{cc}
	\includegraphics[width=3.1in]{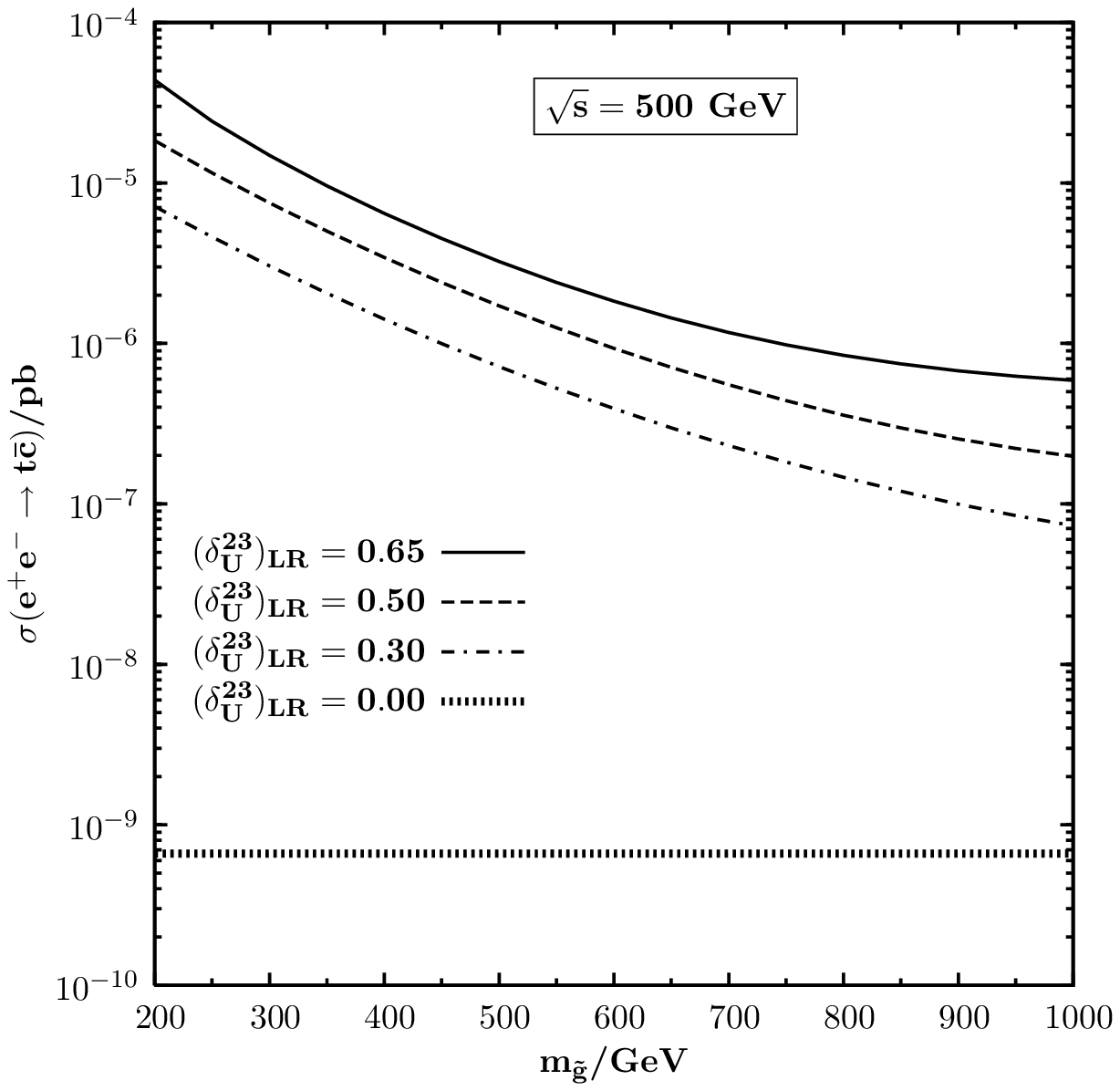} &
	\includegraphics[width=3.1in]{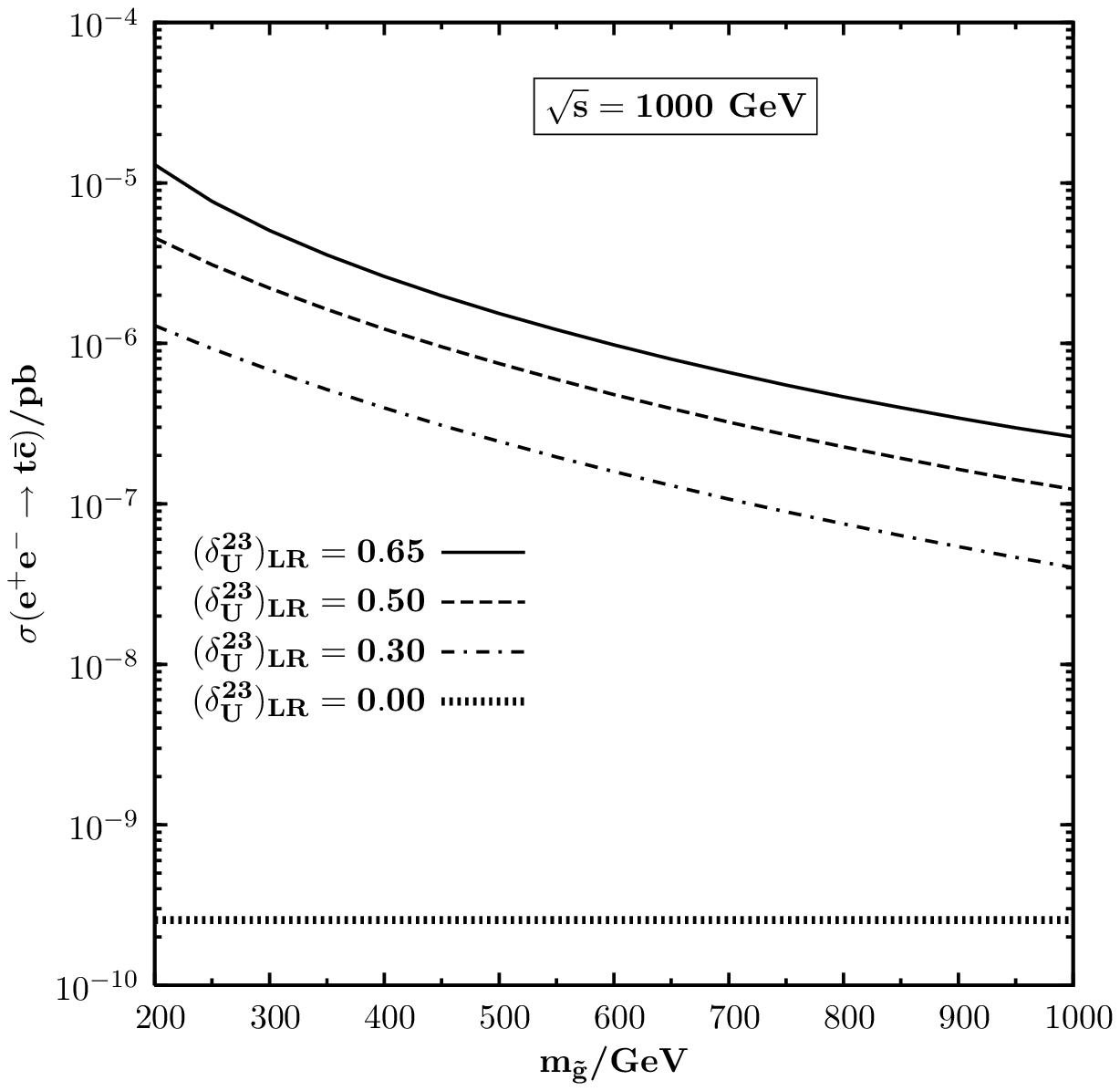} \\
	\end{array}$
\end{center}
\vskip -0.2in
\caption{
\underline{Left panel}:
The cross section of $e^+e^-\to t\bar{c}+{\bar t}c$ as 
a function of $m_{\tilde{g}}$ at $\sqrt{s}=500$ GeV.
\underline{Right panel}:
The cross section as 
a function of $m_{\tilde{g}}$ at $\sqrt{s}=1000$ GeV. 
The parameters are chosen as $\tan\beta=10, M_2=\mu=200$ GeV, 
$M_{\rm SUSY}=250$ GeV, and $A_t=400$ GeV.  
}\label{fig:CS_bst}
\end{figure}

The Feynman diagrams contributing to  $e^+e^- \to t {\bar c}$ process in MSSM can be easily read from the diagrams presented for the decay $t\to c l^+l^-$ in the previous sections. So, we don't repeat them here. Throughout this section the final state of the process is represented as $t\bar{c}$ even though we symmetrize the final state by adding the charged conjugated part as well ($t\bar{c}$ +$\bar{t} c$). The ultra-violet convergence of the process has been checked not only numerically but also analytically. After taking into account of the unitarity properties of the CKM matrix, the $6\times 6$ matrix $\Gamma_U$ in the gluino sector, the $2\times 2$ matrices $U$ and $V$ in the chargino sector, and $4\times 4$ matrix $N$ in the neutralino sector, we have shown analytically that the amplitude of the process is ultra-violet divergent free. A similar analysis was carried out for the decay $t\to cl^+l^-$ discussed in the previous sections. In this section we adopt the same experimental bounds as in $t\to c l^+l^-$ decay. We also apply an angular cut ($10^\circ$) in the center of mass frame. In addition to the cross section, the forward-backward asymmetry ($A_{\rm FB}$)
\begin{equation}
A_{FB}=\frac{\sigma(\theta <90^\circ)-\sigma(\theta >90^\circ)}{\sigma(\theta <90^\circ)+\sigma(\theta >90^\circ)}
\end{equation}
 in $e^+e^- \to t {\bar c}$ as observable is calculated.    

In Fig. \ref{fig:CS}, the cross section $\sigma(e^+e^- \to t {\bar c})$ is presented in two-dimensional contour plots for various parameters. In the upper diagrams, $\sigma(e^+e^- \to t {\bar c})$ in $((\delta_U^{23})_{LR}, \sqrt{s})$ plane is shown with and without GUT relations. The cross section can reach $10^{-5}$ pb only for large flavor-violating parameter values. The enhancement is about 3-4 orders of magnitude with respect to the unconstrainted MSSM results. In the lower diagrams, the contour diagram in $(M_{\rm SUSY}, A_t)$ shows that a small $M_{\rm SUSY}$ with $A_t>500$ GeV favors a cross section of the order of $10^{-5}$ pb. The contours in $((\delta_U^{23})_{LR},(\delta_U^{23})_{LL})$ plane show that a large cross section can be achieved by either assuming only one or both of the flavor-violating parameters large. The cross section can still get as large as $10^{-5}$ pb. We further analyzed the dependence of the cross section with $\tan\beta$ and $M_2$  and found no significant change. 
   
In Fig. \ref{fig:CS_bst} we present a best case scenario for intermediate $\tan\beta$ values as a function of the gluino mass $m_{\tilde{g}}$ at the center of mass energies 500 and 1000 GeV, respectively. The cross section can become only few times $10^{-5}$ pb for a very light gluino and is less than $10^{-9}$ pb for the constrained MSSM case.
\begin{figure}[htb]
\begin{center}$
	\begin{array}{cc}
\hspace*{-0.5cm}
	\includegraphics[width=3.1in]{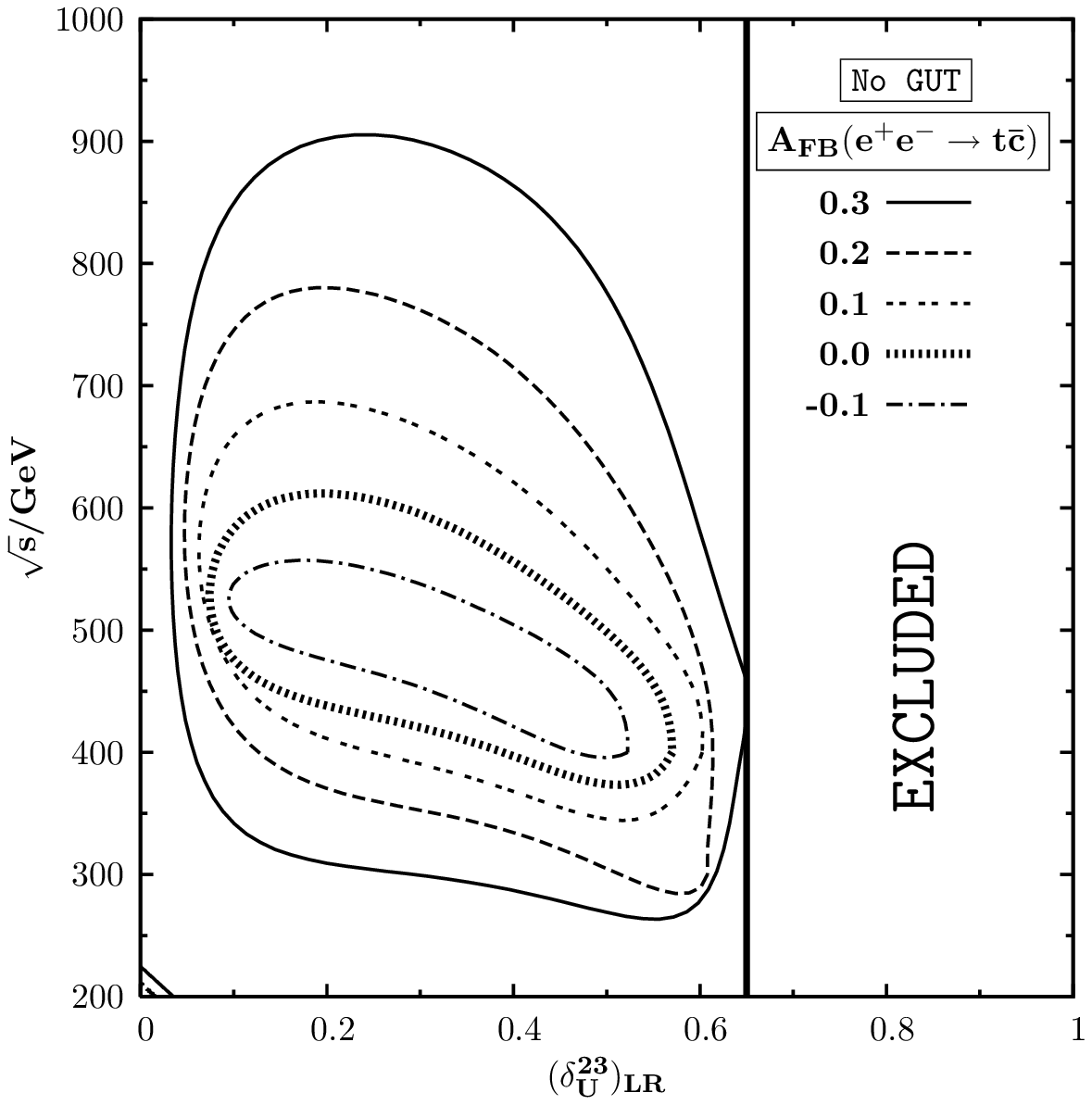} &
	\includegraphics[width=3.1in]{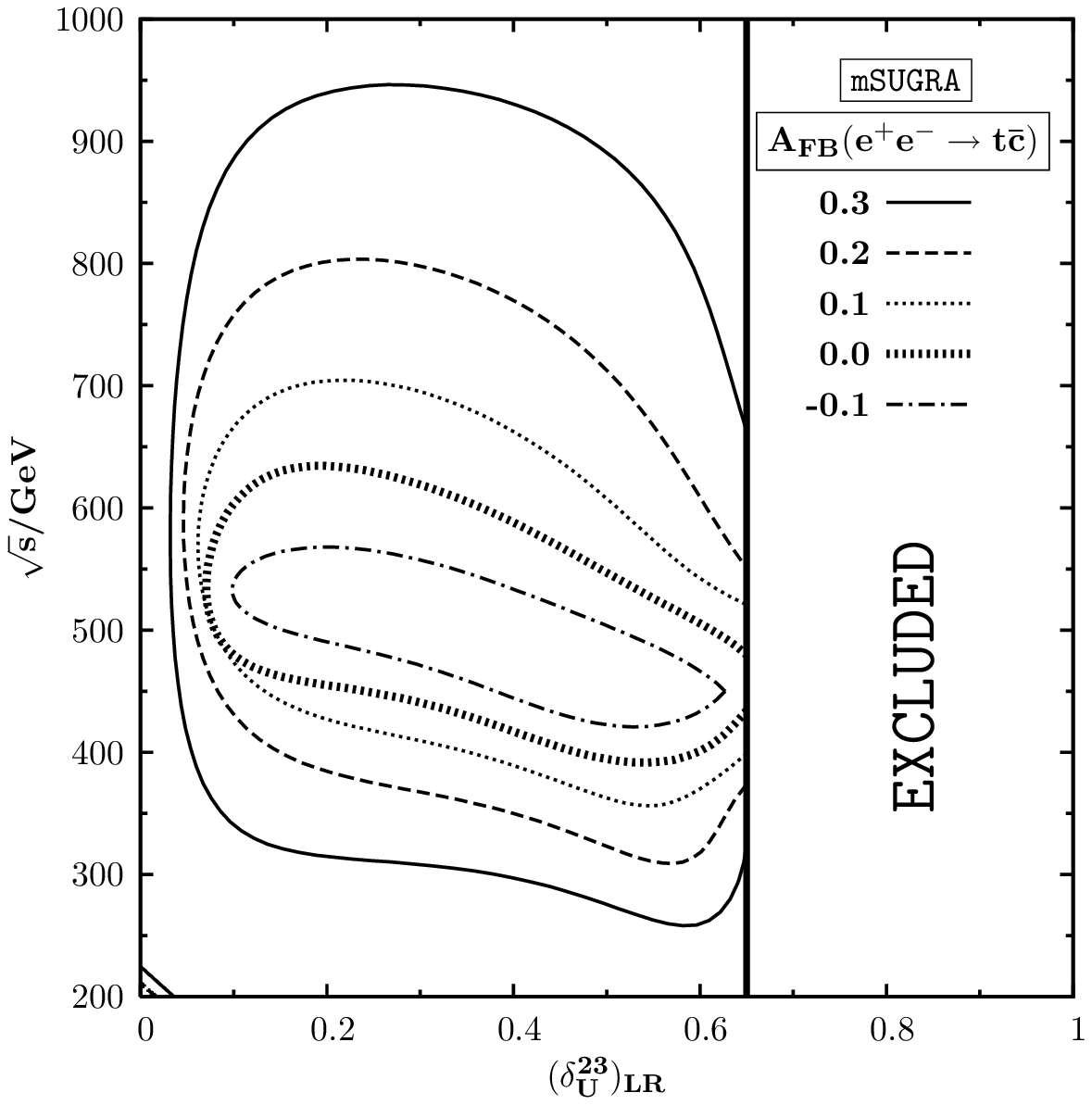} \\
\hspace*{-0.5cm}
	\includegraphics[width=3.1in]{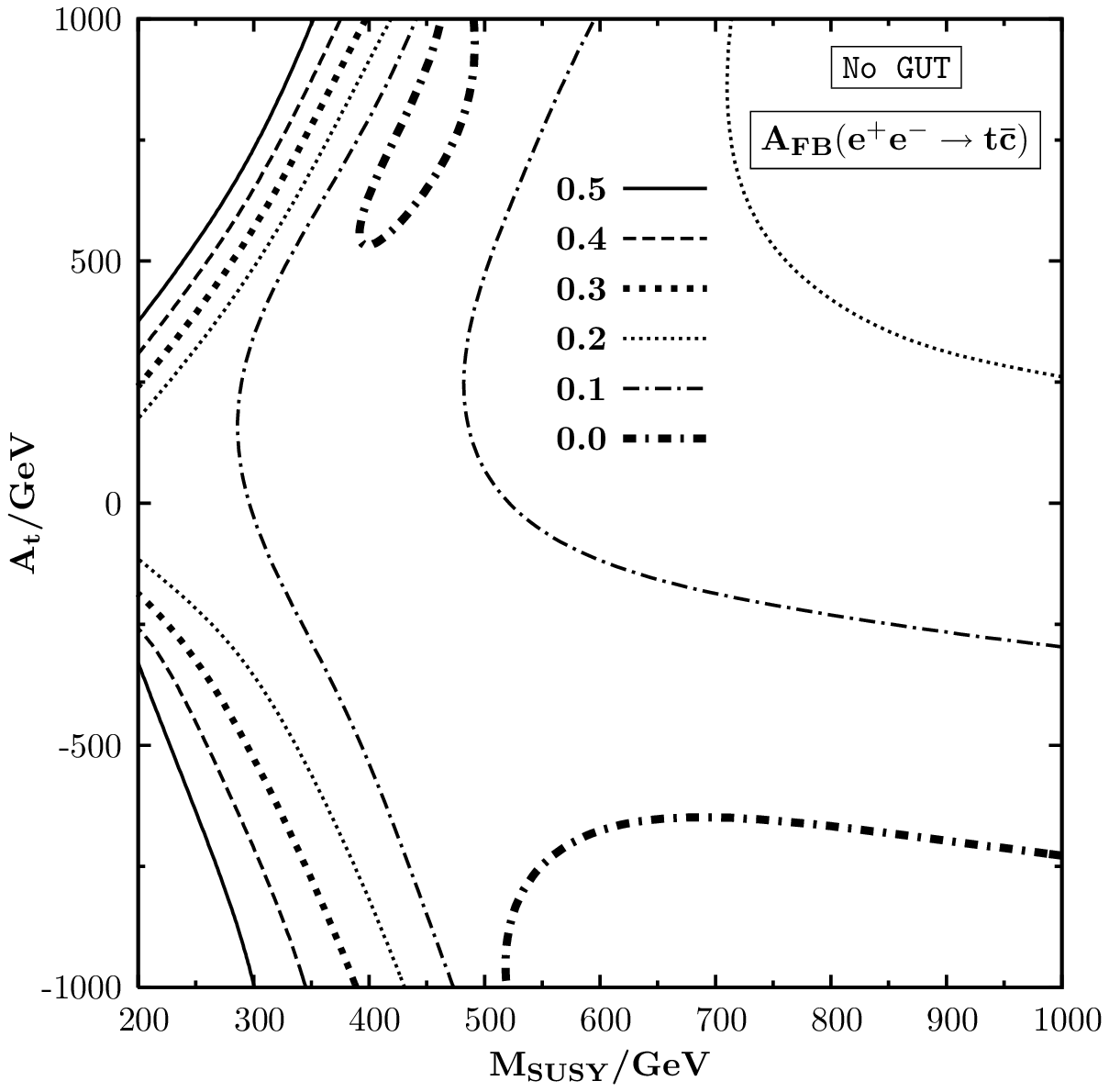} &
	\includegraphics[width=3.1in]{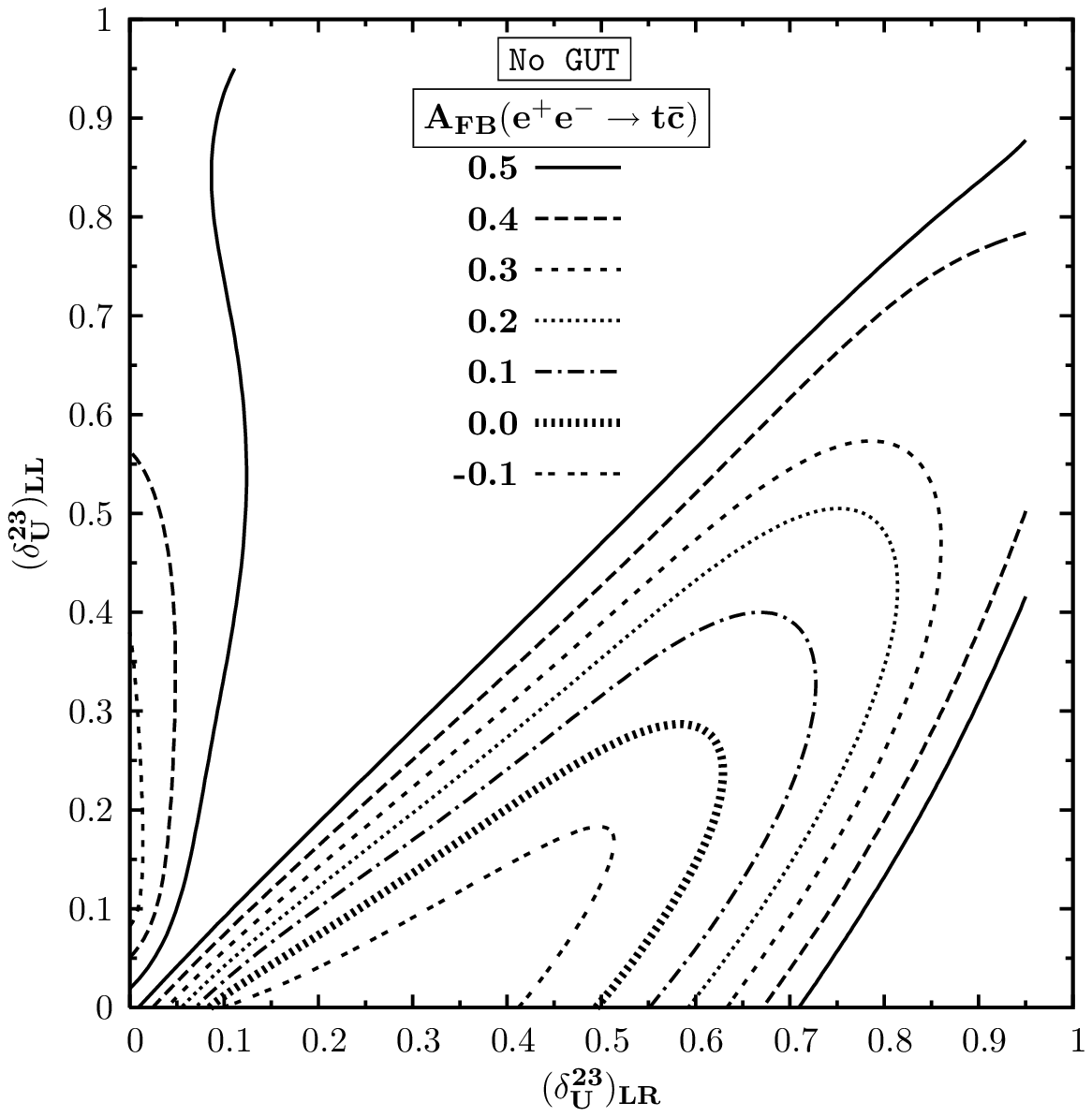} \\

	\end{array}$
\end{center}
\vskip -0.2in
\caption{
\underline{Upper left panel}:
The forward-backward asyymetry, $A_{FB}$, for the process
$e^+e^-\to t\bar{c}$ in 2-dimensional contour plot  
without the
GUT relations ($m_{\tilde{g}} =250$ GeV).
\underline{Upper right panel}:
$A_{FB}$ in 2-dimensional contour plot
in mSUGRA scenario.
The parameters are chosen as $\tan\beta=10$,
$m_{A^0}=M_{\rm {SUSY}}=500$
GeV, $M_2=\mu=200$ GeV, and $A_t=1.2$ TeV.
\underline{Lower left panel}:
$A_{FB}$ in 
2-dimensional contour plot
without GUT relations ($m_{\tilde{g}} =250$ GeV) with 
$(\delta_U^{23})_{LR}=0.5$.
\underline{Lower right panel}:
$A_{FB}$ in 
2-dimensional contour plot
without GUT relations ($m_{\tilde{g}} =250$ GeV).
}\label{fig:AFB}
\end{figure}

The last figure of the section, Fig. \ref{fig:AFB}, shows how the forward-backward asymmetry changes as a function of various MSSM parameters. The upper two diagrams show that the asymmetry $A_{\rm FB}$ can be  as large as $30\%$ for moderate values of $(\delta_U^{23})_{LR}$
at center of mass energies less than $t\bar{t}$ threshold. It is also seen that $A_{\rm FB}$ is not too sensitive to the gluino mass. As shown in the lower diagrams, $A_{\rm FB}$ can become even larger if $M_{\rm SUSY}<300$ GeV with $|A_t|>500$ GeV. For $M_{\rm SUSY}>500$ GeV, the asymmetry is always smaller than $20\%$, independent of the value of $A_t$. Note that the sign of $A_{\rm FB}$ shows the differences between right-handed versus left-handed couplings.      

We would like to end this section by commenting on the observibility of both the decay $t\to c l^+l^-$ and the production $e^+e^-\to t\bar{c}+\bar{t}c$ at ILC. In the case where the center of mass energy is less than $2 m_t\sim 350$ GeV, we shouldn't expect any $t\bar{t}$ background so that we can assume the signal as top quark plus almost a massless c-jet. If we further assume that the integrated luminosity $500 fb^{-1}$ is reached, one should expect to have more than 10 events at $\sqrt{s}=300$ GeV, $m_{\tilde{g}}=250$ GeV with a large flavor violation $(\sigma \sim 0.025 fb)$. This could still be an observable signal after including detector efficiency factors since the  background is very clear. If the center of mass energy is higher than the $2m_t$ threshold, then background cuts will be required due to the $t\bar{t}$ production and the situation becomes worse than in the below threshold case. For the decay $t\to cl^+l^-$, ILC might not be the best place to look. Assuming $\sqrt{s}< 2 m_t$, one can consider $e^+e^-\to t\bar{c}\to c{\bar c}l^+l^-$ where $l$ should be taken either as electron or muon. Due to its short life-time, $l=\tau$ case is much more challenging. Without doubt, this wouldn't give any observable signal (considering the two top channel not feasible). For $\sqrt{s}>500$ GeV, $e^+e^-\to t\bar{t}\to \bar{b}Wcl^+l^-$, which yields more events but a more complicated background. For 
$\sigma(e^+e^-\to t\bar{t})\sim 1$ pb, the number of events for the decay at integrated luminosity $500 fb^{-1}$ becomes around $1.1\times 10^{5}\times BR(t\to cl^+l^-)$.\footnote{W boson is assumed to decay leptonically. In this case, further cuts are necessary to distinguish the lepton from W decay with the signal.} Since $BR(t\to cl^+l^-)\,, l=e,\mu$ can reach at most $10^{-6}$, we  get less than one event. Further cuts needed will make $t\to cl^+l^-$ not possible to be observed. However, if one looked at this decay at LHC, one can have $1.8\times 10^7\times BR(t\to cl^+l^-)$ number of events at $100 fb^{-1}$ integrated limunosity. A total efficiency of $10\%$ is enough to get roughly 1.8 events.

\clearpage
\section{Summary and Conclusion}

In the next few years, combined studies from the hadron (LHC) and linear (ILC) colliders, with their ability to produce large number of top quarks, should be able to test FCNC in its decays. Nonexistent at tree level, and very suppressed at one loop level in the SM, a signal of these processes will most certainly be a sign of New Physics. The dominant and most studied of these decays are the penguin ones: $t \to c \gamma, cg, cZ$ and $cH$. But a complete study of rare decays is likely to unravel more about FCNC phenomenology. We have concentrated here on the rare decays $t \to c l^+ l^-$ and $t \to c \nu_i{\bar \nu}_i$. While the dominant contribution to these decays comes from the penguin diagrams, followed by the decays $\gamma, Z, H \to l^+l^-$, these decays can proceed even in the case in which the two-body decays are forbidden, through box diagrams. Additionally, one can study the associated production cross section at ILC, $e^+e^- \to t {\bar c}$, which is likely to provide a striking and almost background-free signal for single top quark production.

Evaluation of the branching ratio for the decays $t \to c l^+ l^-$ and $t \to c \nu_i{\bar \nu}_i$ has shown that they are strongly suppressed in SM and 2HDM; the branching ratios are expected to be of ${\cal O}(10^{-15})$ for $e^+e^-$ and $\mu^+ \mu^-$ and ${\cal O}(10^{-14})$ for $\tau^+\tau^-$ and $\nu_i {\bar  \nu}_i$.  These values, while small, are of the same order of magnitude, or only one order smaller, than the two body decays. Though at this level neither are possible to observe at present or future colliders, it gives credence to the study of the rare decays $t \to c l^+ l^-$ and $t \to c \nu_i{\bar \nu}_i$ in beyond SM scenarios. In the constrained MSSM the branching ratios are of the same order, or even smaller than in the SM. However, in the unconstrained MSSM branching ratios of ${\cal O}(10^{-7})$ for $e^+e^-$ and $\mu^+ \mu^-$ and ${\cal O}(10^{-6})$ for $\tau^+\tau^-$ and $\nu_i {\bar  \nu}_i$ can be obtained in the mSUGRA scenario; and these branching ratios can be one order of magnitude larger if one relaxes the GUT requirement on gaugino masses. The values obtained are comparable to the values for branching ratios obtained for $t \to c \gamma, cZ$ and $cH$ \cite{reviews}.

We also investigated the single top quark production in $e^+e^- \to t {\bar c}+{\bar t}c$ in the MSSM with and without GUT scenarios. The cross section can get as large as a few times $10^{-5}$ pb for the light gluino case with large flavor violation at $\sqrt{s}= 500$ GeV. This represents more than 5 orders of magnitude enhancement with respect to unconstrained MSSM prediction $(\delta's=0)$. In addition to the cross section, we calculated the forward-backward asymmetry  and found large asymmetries in certain parts of the parameter space (it can become $50\%$) so that such a measurement would be a better alternative to measuring cross sections for observing this channel. We commented on the observibility of both the decay and the production processes at the ILC. Lower energies (less than $t\bar{t}$ threshold) are favorable for the process $e^+e^- \to t {\bar c}$ and we predicted roughly more than 10 events for certain parameter values. At energies larger than the $t\bar{t}$ threshold, the background becomes more challenging. For the decay $t\to cl^+l^-\,, l=e,\mu$, LHC would be a better alternative for observing a signal than ILC, and could give an observable event rate if the total efficiency is around $10\%$.


\end{document}